\documentclass[12pt]{article} 
\pdfoutput=1
\usepackage{amssymb,graphicx}
\usepackage{epstopdf}
\usepackage{amsmath,amsfonts}
\usepackage{epsfig} 
\usepackage{graphicx,graphics}
\usepackage{xcolor}

\begin{document}

\sloppy

\title{\bf Observing primordial magnetic fields through Dark Matter}
\author{Sabir~Ramazanov, Federico~R.~Urban, and Alexander~Vikman\\
\small{\em CEICO, Institute of Physics of the Czech Academy of Sciences,}\\
\small{\em Na Slovance 1999/2, 182 21 Prague 8, Czech Republic}
}

{\let\newpage\relax\maketitle}

\begin{abstract}
Primordial magnetic fields are often thought to be the early Universe seeds that have bloomed into what we observe today as galactic and extra-galactic magnetic fields. Owing to their minuscule strength, primordial magnetic fields are very hard to detect in cosmological and astrophysical observations. We show how this changes if a part of neutral Dark Matter has a magnetic susceptibility. In this way, by studying Dark Matter one can obtain information about the properties of primordial magnetic fields, even if the latter have a comoving amplitude $B_0 \lesssim0.01~\mbox{nG}$. In our model Dark Matter is a stable singlet scalar $\chi$, which interacts with electromagnetism through the Rayleigh operator as $\chi^2 F_{\mu \nu} F^{\mu \nu}/\Lambda^2$. For primordial magnetic fields present in the early Universe this operator forces the $Z_2$-symmetry of the model to be spontaneously broken. Later, when the primordial magnetic field redshifts below a critical value,  the symmetry is restored through an ``inverse phase transition''. At that point the field $\chi$ begins to oscillate and acts as a ``magnetomorphic'' Dark Matter component, inheriting the properties of the primordial magnetic field space distribution. In particular, for a nearly flat spectrum of magnetic field fluctuations, the scalar $\chi$ carries a \emph{statistically anisotropic isocurvature mode}. We discuss the parameter space of the model and consider the possibility that the bulk of the Dark Matter is composed of the same particles $\chi$ produced via the freeze-in mechanism.
\end{abstract}

\section{Introduction} 

The origin and even the existence of a primordial magnetic field (PMF) is debated in the literature~\cite{Durrer:2013pga, Subramanian:2015lua, Giovannini:2003yn}. The main argument in its favour comes from the observation of magnetic fields in galaxies~\cite{Beck:2000dc} including high-redshift galaxies~\cite{Bernet:2008qp}, and clusters~\cite{Carilli:2001hj, Govoni:2004as}. These magnetic fields could indeed be generated by the amplification of the PMF through the dynamo mechanism~\cite{Widrow:2002ud, Kulsrud:2007an}. On the other hand, the same goal could be achieved by using purely astrophysical resources~\cite{Brandenburg:2004jv}. The problem of discriminating between cosmological and astrophysical mechanisms is exacerbated by the fact that a very weak PMF with comoving amplitude $B_0 \simeq 0.01~\mbox{nG}$ or less is sufficient to explain the $B \simeq 1-10~\mu \mbox{G}$ magnetic fields observed in galaxies today~\cite{Dolag:2002bw, Dolag:2004kp}, while also being extremely hard to detect in present and even future cosmological observations. Nonetheless, as we show in this paper, a PMF with strength $B_0 \lesssim 0.01~\mbox{nG}$ is in fact capable of producing visible imprints on the matter distribution in the Universe, provided that it is in contact with another elusive ingredient of the Universe: Dark Matter (DM).

The PMF could be generated either during phase transitions or even before the hot Big Bang. The inflationary mechanisms~\cite{Turner:1987bw, Ratra:1991bn, Martin:2007ue} typically predict a nearly flat spectrum and a correlation wavelength exceeding the present size of the Hubble horizon\footnote{For magnetogenesis in alternative very early Universe scenarios see, e.g., Refs.~\cite{Battefeld:2004cd}--\cite{Frion:2020bxc}.}. Many of these mechanisms often run into strong coupling and backreaction issues~\cite{Demozzi:2009fu, Kanno:2009ei, Urban:2011bu, Emami:2009vd} but several attempts have been made, some successfully, to overcome these problems~\cite{Ferreira:2013sqa}-\cite{Patel:2019isj}. On the other hand, a PMF with a blue spectrum can be generated at much later times during early Universe phase transitions~\cite{Vachaspati:1991nm, Enqvist:1993np, Joyce:1997uy, Durrer:2003ja, Kahniashvili:2012uj}. In that case, the typical comoving correlation wavelength is well below the size of a galaxy, but its power can be transferred to larger scales through the so-called inverse cascade mechanism.

In this paper we do not contribute to the debate on the PMF generation mechanisms. We assume that the PMF exists in the primordial plasma, and that its Fourier modes ${\bf B} ({\bf k}, t)=\int d{\bf x} \, {\bf B} ({\bf x}, t) e^{i{\bf kx}}$ are described by a Gaussian distribution with a two-point correlation function given by 
\begin{equation}
\label{magneticpower}
\langle B_i ({\bf k}, t) B_j ({\bf q}, t)\rangle =(2\pi)^3 \cdot \delta ({\bf k}+{\bf q})  \left(\delta_{ij}-\hat{k}_i \hat{k}_j\right) P_B (k, t) \; .
\end{equation}
Here $P_B (k, t)$ is the power spectrum of the PMF and $\hat{k}_i={k}_i/\left|\mathbf{k}\right|$. It is typically parameterised by a power law, at least for cosmologically interesting wavelengths:
\begin{align}
\label{spectrum}
\begin{cases}
P_{B}(k,t)=A_{B}(t)\cdot k^{n_{B}}\,\quad & \text{for}\quad k<k_{D}(t) \\
P_{B}(k,t)=0 \quad &\text{for}\quad k\geq k_{D}(t)
\end{cases}
\,,
\end{align}
where $n_B$ is the spectral tilt, and $k_D$ is the cutoff scale above which magnetic fields are dissipated in the plasma~\cite{Jedamzik:1996wp, Subramanian:1997gi}. Magnetic fields with momenta below $k_D$ are frozen into the highly conductive primordial plasma, that is, they decay as $B(t) \propto 1/a^2(t)$, whence $A_B (t) \propto 1/a^4 (t)$. The power spectrum of the form~\eqref{magneticpower} describes a non-helical PMF. Adding helicity, which can be generated, e.g., during inflation~\cite{Caprini:2014mja}, is irrelevant for the present work (as we will show later), and we can safely neglect it.

Searches for signatures of the PMF in astrophysical and cosmological data lead to a series of limits on the comoving amplitude $B_0$. The most stringent model-independent upper limit on the PMF integrated over all scales has been recently imposed from the study of its effect on recombination: $B_0 \lesssim 0.05~\mbox{nG}$ and $B_0 \lesssim 0.01~\mbox{nG}$ for $n_B =-3$ and for $n_B =2$ in Eq.~\eqref{spectrum}, respectively~\cite{Jedamzik:2018itu}. Notably, provided that the PMF saturates these bounds, there is an intriguing opportunity to alleviate the tension between the low- and high-redshift measurements of the Hubble constant~\cite{Jedamzik:2020krr}. Another set of limits on the PMF has been obtained from the non-observation of its signatures in the Cosmic Microwave Background (CMB) temperature fluctuations as observed by the Planck collaboration~\cite{Ade:2015cva}. Specifically, these limits follow from the study of the angular power spectrum, searches for magnetically induced non-Gaussianity, preferred directions in the bispectrum, and the Faraday rotation of the CMB polarisation. The most stringent Planck bound on the PMF reads $B_0 \lesssim 1~\mbox{nG}$ at a scale of $1~\mbox{Mpc}$. It is argued in Ref.~\cite{Trivedi:2013wqa} that the PMF may leave a strong imprint on the CMB trispectrum, unless $B_0 \lesssim 0.05~\mbox{nG}$, but this is a model-dependent prediction. The upper limits cited above assume that the PMF exists in the cosmic plasma (well) before recombination. Even if this is not the case, a later PMF would still have an effect on the Faraday rotation measures of distant quasars~\cite{Blasi:1999hu, Pshirkov:2015tua}. Using these data, the resulting upper bound on a PMF with correlation wavelength exceeding the present Hubble radius is given by $B_0 \lesssim 1~\mbox{nG}$. Finally, we would like to mention a tentative {\it lower} bound on the PMF obtained from the non-observation of GeV cosmic rays produced in the cosmic ray cascade triggered by TeV blazars or active galactic nuclei: $B_0 \gtrsim 3 \cdot 10^{-7}~\mbox{nG}$ at a scale of $1~\mbox{Mpc}$~\cite{Neronov:1900zz}. This may serve as another argument in favour of the existence of a PMF, because there is no known astrophysical mechanism that could eject magnetic fields into the extremely low-density inter-galactic medium or even in voids.

To summarise: present data do not allow to probe magnetic fields with strength $B_0 \lesssim 0.01~\mbox{nG}$, which however are those best motivated from the viewpoint of seeding galactic magnetic fields. The main goal of the present work is to show that such a weak PMF could still leave distinct imprints on current cosmological datasets. This is achieved at the price of extending the Standard Model (SM), which should be extended anyway, at least because the SM does not accommodate a suitable candidate for DM. We discuss the possibility of enhancing the visibility of both the PMF and DM by bringing them in direct contact with each other, in such a way that the PMF is responsible for the generation of a fraction of DM\footnote{The opposite possibility, the generation of a PMF from DM, has been entertained in Ref.~\cite{Nugaev:2020zcv}.}. Thus, studying the properties of DM, one could extract information about the PMF.

The model we base our discussion on is presented in Ref.~\cite{Babichev:2020xeg}, which deals with a novel mechanism of particle production in the early Universe. The main ingredient of that mechanism is a singlet scalar $\chi$ equipped with the potential: 
\begin{equation}
V= \frac{\left(M^2 -\mu^2(x)\right) \cdot \chi^2}{2}+\frac{\lambda \chi^4}{4} \; ,
\end{equation}
where $\mu(x)$ is an external field monotonically decreasing as the Universe expands. The field $\mu(x)$ is assumed to be large at early times, so that the $Z_2$-symmetry of the model is spontaneously broken. Later on, the external field $\mu(x)$ redshifts considerably, the bare mass $M$ of the field $\chi$ becomes dominant, and the latter begins to oscillate around zero. This situation is reminiscent of the conventional phase transitions in the early Universe, but with the sequence of events reversed. That is, in standard phase transitions the system is initially in the unbroken phase, and its symmetry is spontaneously broken as the Universe expands and the temperature drops below some critical value. Hence, in the case at hand we can speak of an ``inverse phase transition'', because the symmetry, initially broken, is restored at later times.


In Ref.~\cite{Babichev:2020xeg}, it has been assumed that $\mu^2 (x) \propto R$, where $R$ is the Ricci scalar\footnote{Other realisations of spontaneous symmetry breaking caused by the coupling to the Ricci scalar have been considered in Refs.~\cite{Laulumaa:2020pqi, Bettoni:2018pbl}.}. Note, however, that any other cosmological field monotonically decreasing with time could be considered as the driver of the inverse phase transition and the consequent $\chi$ oscillations. In the present work we explore the possibility that the function $\mu^2(x)$ is sourced by the PMF. This is realised with the choice
\begin{equation}
 \mu^2(x) \propto F_{\mu \nu} F^{\mu \nu} \; ,
\end{equation}
where $F_{\mu \nu}$ is the electromagnetic field strength. Then, given that $F_{\mu \nu} F^{\mu \nu}=2\left(B^2-E^2 \right)$, the interaction between the field $\chi$ and the PMF ${\bf B}$ is simply $\chi^2 B^2$. Notice that the electric field ${\bf E}$ is completely negligible because of the high conductivity of the primordial plasma~\cite{Hollenstein:2012mb}.

The field $\chi$, if it is stable on cosmological timescales, contributes to DM. Remarkably, even a very weak PMF with comoving amplitude $B_0 \lesssim 0.01~\mbox{nG}$ is enough to generate a considerable fraction of DM in the form of the field $\chi$, which we dub ``magnetomorphic DM''. Therefore, by studying the properties of the DM field $\chi$ one can infer the properties of the fluctuations of the PMF. For the purpose of demonstrating the viability of the mechanism, in this paper we focus on the tractable example of a PMF with a nearly scale-invariant power spectrum. In this case the PMF can be naturally split into a background part composed of very long wavelength modes, and shorter scale modes which are interpreted as fluctuations. The most notable feature of this model is that \emph{the isocurvature perturbations of the field $\chi$ are statistically anisotropic}, that is, their power spectrum is direction-dependent, cf.\ Refs.~\cite{Nakayama:2019rhg, Nakayama:2020rka}. This then becomes the main prediction of our mechanism.

The outline of the paper is as follows. In Section~2 we review the basic dynamics of the field $\chi$ during the inverse phase transition and calculate its energy density. In Section~3 we focus on the situation in which the spontaneous symmetry breaking is due to the PMF. In Section~4 we show how the superhorizon fluctuations of the PMF leave an imprint on DM fluctuations. A viable range of parameter space is discussed in Section~5. We conclude in Section~6 with a discussion of our results.

\section{The inverse phase transition}

In the present work we are interested in the dynamics of a singlet scalar field, $\chi$, which is described by the general Lagrangian~\cite{Babichev:2020xeg}
\begin{equation}
{\cal L}= \frac{(\partial \chi)^2}{2} -\frac{\left(M^2-\mu^2 (t, {\bf x})\right) \cdot \chi^2}{2}-\frac{\lambda \chi^4}{4} \; .
\end{equation}
Here $\mu(t, {\bf x})$ is an external field, assumed to be changing only due to cosmic 
expansion and to be nearly homogeneous, i.e., $\mu^2 (t, {\bf x}) \approx \mu^2 (t)$. For initially large $\mu^2 (t)$, spontaneous symmetry breaking may take place at very early times. Namely, the field $\chi$, initially in the broken symmetry phase, resides in the potential minimum, which slowly changes following the redshift of the function $\mu^2 (t)$:
\begin{equation}
\label{vacuum}
\chi (t)= \sqrt{\frac{\mu^2 (t)-M^2}{\lambda}} \; .
\end{equation}
This is true provided that the effective mass squared given by
\begin{equation}
\label{def}
M^2_{eff} =3\lambda \chi^2+M^2 -\mu^2(t) \; ,
\end{equation}
slowly changes on time scales of order of $M^{-1}_{eff}$:
\begin{equation}
\label{ad}
\left|\frac{\dot{M}_{eff}}{M^2_{eff}} \right|\ll 1 \; .
\end{equation}
Substituting Eq.~\eqref{vacuum} into Eq.~\eqref{def}, we find
\begin{equation}
M^2_{eff} \left(t\right)=2 \cdot ( \mu^2 (t)-M^2 ) \; .
\end{equation}
Using this in the inequality~\eqref{ad} we obtain
\begin{equation}
\label{ineq}
\kappa H \mu^2 (t) \ll 2\sqrt{2} \cdot \left(\mu^2 (t)-M^2 \right)^{3/2} \; ,
\end{equation}
where $\kappa$ is defined from
\begin{equation}
\label{kappadef}
\frac{d \mu^2 (t)}{dt}=\kappa H \mu^2 (t)\; .
\end{equation}
Because we assume that the external field $\mu(t, {\bf x})$ varies only following the Hubble drag, typically one has $\kappa \simeq {\cal O} (1)$. As soon as the condition~\eqref{ineq} is violated at some time $t=t_*$, the field $\chi$ stops to track the minimum of the potential, and starts oscillating. The initial amplitude of the oscillations $\chi_*$ follows from Eq.~\eqref{vacuum}, which we extrapolate to the transition time $t \simeq t_*$, and Eq.~\eqref{ineq} replaced by the approximate equality:
\begin{equation}
\label{equality} 
\chi_* \simeq \frac{(\kappa H_* \mu^2_*)^{1/3}}{\sqrt{2\lambda}} \; .
\end{equation}
Hereafter the subscript `$*$' indicates that the corresponding quantities are taken at the time $t_*$. The evolution of the system corresponds to a phase transition in the early Universe. Contrary to conventional cases, however, the field $\chi$ is initially in the spontaneously broken phase, which is restored at later times. Thus, we call this mechanism the inverse phase transition.

There are two distinct situations for which the condition~\eqref{ineq} is no longer valid. The first one is when $\mu(t)$ drops faster than $H(t)$ and at the transition we have $\mu^2_* \gg M^2$ while $\mu_*\sim \kappa H_*$. We discuss this possibility at the end of the Section for completeness. However, in this paper we are primarily interested in the other option, that is, when the external field $\mu(t)$ decreases at the same rate as the Hubble parameter, so that $\mu(t) \gg H(t)$ at all relevant times. In that case, the inequality~\eqref{ineq} is violated as $\mu(t)$ approaches $M$ at the time $t_*$:
\begin{equation}
\label{equalitygen}
\mu^2_* \approx M^2 \; .
\end{equation}
Hence, at the time $t_*$, the mass $M$ is larger than the Hubble rate: 
\begin{equation}
\label{hierarchy}
M \gg H_* \; .
\end{equation}
Combining Eqs.~\eqref{equality} and~\eqref{equalitygen}, we rewrite the amplitude $\chi_*$ as
\begin{equation}
\label{maincase}
\chi_* \simeq \left(\frac{\kappa^2}{8} \right)^{1/6} \cdot \frac{M^{2/3} H^{1/3}_*}{\sqrt{\lambda}} \; .
\end{equation}
The resulting energy density stored by the oscillating field $\chi$ is given by
\begin{equation}
\label{DMenergy}
\rho_{\chi} (t)=\frac{M^2 \chi^2_*}{2} \cdot \left(\frac{a_*}{a(t)} \right)^3 \simeq  \frac{(\kappa \cdot M^5 \cdot H_*)^{2/3}}{ 4\lambda} \cdot \left(\frac{a_*}{a(t)} \right)^3 \; .
\end{equation}
Notably, the energy density $\rho_{\chi} (t)$ non-perturbatively depends on the coupling constant $\lambda$. This should not be surprising, because the inverse phase transition is an intrinsically non-perturbative process. The physical interpretation of this dependence on $\lambda$ is clear: for smaller $\lambda$ the initial offset of the field $\chi$ is larger, and so is the initial amplitude of oscillations.

The inhomogeneities of the external field $\mu(t, {\bf x})$ result into the inhomogeneities of the time $t_* ({\bf x})$ at which the symmetry is restored. Consequently, at different points in space the field $\chi$ starts oscillating at slightly different times. This produces slight inhomogeneities in the energy density $\rho_{\chi}$. As a result, the field $\chi$ carries information about the primordial fluctuations of the external field $\mu(t, {\bf x})$ . This is the source of the non-trivial phenomenology through which the inverse phase transition could be tested. We discuss this in Section~4, after specifying the origin of the field $\mu (t, {\bf x})$.

Before we close this Section, let us briefly discuss the case when the condition~\eqref{ineq} is violated while $\mu^2_* \gg M^2$, or, equivalently, $M \ll H_*$. This naturally occurs if $\mu(t)$ decreases faster with time than the Hubble rate. In that case, the value of the external field at the time when Eq.~\eqref{ineq} ceases to be valid is estimated to be $\mu_* \simeq \kappa H_* / 2\sqrt{2}$. Using Eq.~\eqref{equality} we then obtain:
\begin{equation}
\chi_* \simeq \frac{\kappa}{2} \cdot \frac{H_*}{\sqrt{2\lambda}} \; .
\end{equation}
After decoupling the field $\chi$ is oscillating in the quartic potential well until the time $t =t_M$ at which its amplitude, decreasing as $1/a$, drops down to 
\begin{equation}
3\lambda \chi^2_{M} \simeq M^2 \; ;
\end{equation}
(the subscript $`M`$ stands for the moment of time $t=t_M$). From this point on, the dynamics of the field $\chi$ is dominated by the quadratic potential. The amplitude $\chi_M$ is given by 
\begin{equation}
\chi_M \simeq \chi_* \cdot \left(\frac{a_*}{a_M} \right) \; .
\end{equation}
Combining the latter two equations, one obtains 
\begin{equation}
\chi_M \simeq \frac{M}{ \sqrt{3\lambda}} \; ,
\end{equation}
and 
\begin{equation}
\frac{a_M}{a_*} \simeq \sqrt{\frac{3}{8}} \cdot \frac{\kappa H_*}{M} \; .
\end{equation}
Finally, we obtain the energy density of the field $\chi$ in the case $M \ll H_*$:
\begin{equation}
\rho_{\chi} (t) \simeq \frac{M^4}{6\lambda} \cdot \left(\frac{a_M}{a(t)} \right)^3 \; ,
\end{equation} 
or, equivalently:
\begin{equation}
\rho_{\chi} (t) \simeq \left(\frac{3\kappa^2}{8} \right)^{3/2} \cdot \frac{M H^3_*}{6\lambda} \cdot \left(\frac{a_*}{a(t)} \right)^3 \; .
\end{equation} 
We can thus conclude that in both cases, when $M \gg H_*$ and $M \ll H_*$, for sufficiently heavy masses $M$ and small $\lambda$ the energy density of the field $\chi$ can be large enough to give a substantial contribution to the total energy budget of the Universe today.

\section{Turning on the primordial magnetic field}

The phenomenology of the inverse phase transition is especially rich when the field $\mu(t, {\bf x})$ is sourced by the PMF. This is achieved by choosing 
\begin{equation}
\label{relation}
\mu^2 =\frac{F_{\mu \nu} F^{\mu \nu}}{2 \Lambda^2} \; .
\end{equation}
That is, we add the leading order, $Z_2$-symmetric non-renormalisable interaction between the field $\chi$ and the electromagnetic field strength $F_{\mu \nu}$, with a UV cutoff scale set by the constant $\Lambda$,
\begin{equation}
\label{nonrenorm}
\frac{\chi^2 F_{\mu \nu} F^{\mu \nu}}{2 \Lambda^2} \;
\end{equation}
(the factor `1/2' is added here for the future convenience). This non-renormalisable operator, the CP-invariant Rayleigh operator, arises naturally in many extensions of the SM, e.g., if the field $\chi$ interacts with some heavy fermions charged under the electromagnetic $U(1)$ group, or if the field $\chi$ interacts with electromagnetism through a scalar mediator as in secluded DM models~\cite{Chang:2010en, Weiner:2012gm, Liu:2013gba, Rajaraman:2012fu, Dudas:2014ixa}. This term is responsible for the magnetic susceptibility  of $\chi$, and appears in the relativistic theory of Casimir-Polder and Van der Waals interactions~\cite{Feinberg1970, Itzyk}. In what follows, we do not assume a particular UV-completion or origin for this interaction.

A comment is in order before we proceed. As it will become clear soon, in our setup the inverse phase transition occurs at very high temperatures $T_* \gg 100~\mbox{GeV}$, when electromagnetic and weak forces are unified, and the gauge sector of the SM is described by the massless vector bosons of the electroweak group $SU (2) \times U(1)$. While this is true for the primordial hot plasma, our version of the inverse phase transition is an intrinsically out-of-equilibrium process which involves very low momenta of the relevant fields, momenta that are well below the temperatures of the early Universe. In this low-momentum regime the gauge sector of the SM effectively reduces to that of electromagnetism.

In terms of the electric and magnetic fields $F_{\mu \nu} F^{\mu \nu}=2 \cdot \left( B^2-E^2 \right)$. Owing to the high conductivity of the primordial plasma throughout the history of the Universe, the electric field ${\bf E}$ can be set to zero. At the same time, we assume that the plasma is permeated by a long-range magnetic field ${\bf B} (t, {\bf x})$. Thus, one can write
\begin{equation}
\label{relationPMF} 
\mu^2 (t, {\bf x})=\frac{B^2 (t, {\bf x})}{\Lambda^2} \; .
\end{equation}
Hereafter we assume that the inverse phase transition takes place during the radiation-dominated stage, unless otherwise stated. In the magnetohydrodynamic limit, which applies in this case, the PMF is frozen-in, and its time-dependence is given by
\begin{equation}
B (t, {\bf x})=B_0 ({\bf x})\cdot \left(\frac{a_0}{a(t)} \right)^2 \; .
\end{equation}
In particular, this means that the quantity $\kappa$ defined by Eq.~\eqref{kappadef} is given by $\kappa =4$.

Before switching to the discussion of the evolution of $\chi$ as caused by the PMF, let us specify the details of the PMF configuration. We assume that the power of the PMF mostly resides at large scales, exceeding the size of the present horizon, as is expected in case of nearly-flat spectrum (see below). In this situation shorter scale modes are naturally interpreted as fluctuations on the approximately homogeneous magnetic background. In other words, one can decompose the PMF as follows:
\begin{equation}
{\bf B} (t, {\bf x}) ={\bf B} (t)+\delta {\bf B} (t, {\bf x}) \; ,
\end{equation}
where $|\delta {\bf B}| \ll B (t)$. In particular, this configuration is natural if the PMF possesses a nearly flat spectrum, i.e., $n_B \approx -3$. As follows from Eq.~\eqref{magneticpower}, the homogeneous part of the PMF is related to the power spectrum by
\begin{equation}
{\bf B}^2 (t) =\int_{\frac{k}{a_0} <H_0} \frac{k^2 \, dk \, P_B (k)}{\pi^2} \propto \ln \frac{H_0}{\Lambda_{IR}}\; .
\end{equation} 
The lower bound $\Lambda_{IR}$ on the comoving wavenumber $k$ depends on the mechanism of PMF generation. For example, for inflationary mechanisms it depends on the number of e-folds beyond the $N_{CMB} \approx 50-70$~\cite{Bartolo:2012sd,Thorsrud:2013mma,Thorsrud:2013kya} at which the cosmological mode with wavelength approximately equal to the present Hubble radius exits the horizon. The physical origin of the homogeneous ${\bf B} (t)$ does not matter for us: it suffices to know that it is generically present.  This allows us to treat the PMF in terms of linear perturbation theory.

In this Section we consider the effect of the homogeneous part of the PMF, ${\bf B} (t)$, and defer the treatment of fluctuations to the next Section. The effect of ${\bf B} (t)$ on the background geometry of the Universe 
is negligible because its energy density is very small; moreover, the homogeneous mode is compensated by the anisotropic pressure of free-streaming neutrinos~\cite{Adamek:2011pr}. Therefore, we can work with isotropic background in what follows. Because the inverse phase transition takes place during the radiation-dominated epoch, both the PMF, and hence $\mu(t)$, as well as the Hubble rate $H(t)$, redshift as $1/a^2(t)$. According to the discussion of the previous Section, in that case the approximate equality~\eqref{equalitygen} must be fulfilled at the transition. With the use of Eq.~\eqref{relationPMF} this gives
\begin{equation} 
M^2 \simeq \frac{B^2_*}{\Lambda^2} \; . 
\end{equation}
It is convenient to work in terms of the magnetic\footnote{In all formulas we use Heaviside-Lorentz units, whereas the numerical values of magnetic field are provided in Gauss.} to radiation energy density ratio $r_B (t)$:
\begin{equation}
\label{rB}
r_B (t) \equiv \frac{\rho_B (t)}{\rho_{rad} (t)} =\frac{B^2}{2\rho_{rad} (t)} \; .
\end{equation}
This quantity is practically constant, and changes only due to the change of ultra-relativistic degrees of freedom $g_* (T)$: 
\begin{equation}
\label{revolution} 
r_B (t_*) =r_B (t_0) \cdot \left( \frac{g_* (T_*)}{g_* (T_0)} \right)^{1/3} \; .
\end{equation}
This follows from the conservation of entropy in a comoving volume $s \cdot a^3 =\mbox{const}$, where $s \approx 2\pi^2 g_* (T) T^3/45$. The energy density of radiation reads
\begin{equation}
\rho_{rad} (T) =\frac{\pi^2 g_* (T) \cdot T^4}{30} \; .
\end{equation}
we then obtain the relation between the mass $M$ and the transition temperature $T_*$:
\begin{equation}
\label{massrB}
M^2 \approx \frac{\pi^2 r_B (t_0)}{15} \cdot \frac{g^{4/3}_* (T_*)}{g^{1/3}_* (T_0)} \cdot \frac{T^4_*}{\Lambda^2} \; .
\end{equation}
Therefore, typically we have $M \ll T_*$. This statement is further strengthened in Section~5 where we discuss several constraints on the parameter space.

Barring fine-tuned situations, the field $\chi$ will either be unstable and decay rapidly, well before the end of radiation-dominated stage, or be stable on timescales of the order of the age of the Universe. In the former case any imprint of the PMF is going to be completely diluted in the background radiation. In the latter, more interesting case, the field $\chi$ constitutes a fraction of DM:
\begin{equation}
\label{ft} 
f \equiv \frac{\rho_{\chi}(t)}{\rho_{DM}(t)} \; .
\end{equation}
Assuming that the rest of DM, be it $\chi$ itself or some other field, is produced before or at $t_*$, from this point on $f$ is a constant quantity, and can be evaluated at any time. We choose to calculate it at the time of matter-radiation equality, $t=t_{eq}$, when 
\begin{equation}
\label{dmradeq}
\rho_{DM, eq} \approx \rho_{rad, eq} \approx \frac{\pi^2 g_* (T_{eq}) T^4_{eq}}{30} \; .
\end{equation}
At equality one has $T_{eq} \approx 1~\mbox{eV}$ and $g_* (T_{eq}) \approx 3.4$. Recall the expression for the energy density of the field $\chi$ given by Eq.~\eqref{DMenergy}, where we take $\kappa=4$. Then, upon using the formula for the Hubble rate during radiation-domination:
\begin{equation}
\label{hubbletemperature}
H (T) =\sqrt{\frac{\pi^2 g_* (T)}{90}} \cdot \frac{T^2}{M_{Pl}} \; , 
\end{equation}
and expressing the temperature from Eq.~\eqref{massrB}, we get the final result for the DM fraction $f$:
\begin{equation}
\label{fraction}
 \frac{f}{0.1} \simeq \left( \frac{10^{-11}}{\lambda} \right) \cdot \left(\frac{10^3}{g_* (T_*)} \right)^{1/9} \cdot \left(\frac{r_B (t_0)}{10^{-10}} \right)^{5/12} \ \cdot \left(\frac{M}{\mbox{TeV}} \right)^{5/2} \cdot \left(\frac{10^{14}~\mbox{GeV}}{\Lambda} \right)^{5/6} \; . 
\end{equation}
For instance, the value $r_B (t_0) \simeq 10^{-10}$ corresponds to $B_0 \simeq 0.03~\mbox{nG}$, whereas $B_0 \simeq 0.01~\mbox{nG}$ gives $r_B (t_0) \simeq 10^{-11}$. Such a weak PMF is interesting because it could be responsible for generating galactic magnetic fields and yet invisible in standard cosmological data. Our choice of normalisation for the other quantities will be clear in what follows. As for the DM fraction $f$, one might wonder whether it is possible to bring that fraction to $f=1$. In most cases the answer is negative, because of the properties of the fluctuations of the field $\chi$, discussed in the next Section, which are drastically different from those of canonical DM; in particular, the field $\chi$ produced via the inverse phase transition carries significant isocurvature power. As a result, the field $\chi$ can constitute only a small fraction $f \ll 1$ of the overall DM (but see Section 5 for a more precise assessment).

\section{The rise of statistically anisotropic isocurvature}

We are primarily interested in a PMF with comoving amplitude $B_0$ well below $1~\mbox{nG}$, the current Planck limit~\cite{Ade:2015cva}. In this case it is safe to neglect any direct effect of PMF fluctuations on metric perturbations calculated, e.g., in Refs.~\cite{Giovannini:2007qn, Yamazaki:2008gr, Paoletti:2008ck, Bonvin:2011dt, Bonvin:2013tba}. Still, the fluctuations of the PMF are not irrelevant, because they source isocurvature perturbations of the field $\chi$, as we will see here. Thus, if the field $\chi$ is stable and contributes to DM, it will also contribute to the isocurvature DM mode, which is strongly constrained by the Planck collaboration \cite{Planck 2018_X}. We focus here on the superhorizon fluctuations of the PMF and their effect on DM.

The isocurvature DM mode is defined by \cite{Malik:2002jb,Kodama:1985bj}
\begin{equation}
\label{iso} 
{\cal S}_{DM}=3 \left(\zeta_{DM}-\zeta_{rad} \right) \; ,
\end{equation} 
where $\zeta_{DM}$ and $\zeta_{rad}$ are gauge-independent curvature perturbations associated with DM and radiation, respectively. Generally, for any $i$-component of the fluid described by the energy density $\rho_i$ and the pressure density ${\cal P}_i$, one can write
\begin{equation}
\zeta_i = \Psi +\frac{\delta \rho_i}{3 (\rho_i+{\cal P}_i)} \; ,
\end{equation}
where $\Psi$ is the gravitational potential which accounts for the perturbations of the $ij$-components of the metric. We assume that $\rho_{\chi} \ll \rho_{DM}$, or equivalently $f\ll1$. This condition should be imposed anyway, because the isocurvature mode seeded by $\chi$ must be kept small. Assuming also that the field $\chi$ is the only source of DM isocurvature perturbations, we obtain
\begin{equation}
\label{isochi}
{\cal S}_{DM}=f \cdot \delta_{\chi, iso} \; .
\end{equation}
Here $\delta_{\chi, iso} \equiv \left(\frac{\delta \rho_{\chi}}{\rho_{\chi}} \right)_{iso}$ is the perturbation $\delta_{\chi}$ sourced by the PMF.

As we show in the Appendix, the relation between the perturbation $\delta_{\chi, iso}$ and the fluctuations of the PMF at the time of the transition $\delta {\bf B}_* ({\bf x})$ is:
\begin{equation}
\label{finaliso}
\delta_{\chi, iso}=\frac{5}{6} \cdot \frac{{\bf B}_{*} \cdot \delta {\bf B}_* ({\bf x})}{B^2_*} \; .
\end{equation} 
That is, the perturbation $\delta_{\chi, iso}$ inherits the properties of the PMF; in other words, PMF perturbations morph into DM ones. Thus, despite the fact that the energy density of the PMF is very small, its fluctuations may become visible if the field $\chi$ constitutes a sizeable fraction of the total DM in the Universe.

The power spectrum of isocurvature perturbations following from Eq.~\eqref{finaliso} is given by 
\begin{equation}
P_{{\cal S}_{DM}}=\frac{25 \cdot f^2}{36}  \cdot \frac{ P_{B_0} (k)}{ B^2_0} \cdot (\hat{{\bf B}} \times \hat{{\bf k}})^2 \; ,
\end{equation}
where $\hat{{\bf B}} \equiv {\bf B}/B$ and $\hat{{\bf k}} \equiv {\bf k}/k$ are the PMF and momentum unit vectors, respectively. This formula highlights the key prediction of this work: \emph{DM isocurvature perturbations have a statistically anisotropic power spectrum}. In other words, their spectrum depends on the direction of the background magnetic field.

The amount of isocurvature perturbations is conventionally quantified by the ratio of isocurvature to adiabatic power spectra, which is strongly constrained by the Planck collaboration: $P_{{\cal S}_{DM}} / P_{\zeta} <0.038$~\cite{Planck 2018_X}. Strictly speaking, statistical anisotropy has been investigated only in the case of adiabatic perturbations. Isocurvature perturbations leave a different imprint on the CMB; thus, their possible direction-dependence should be studied separately. Still, some conclusions regarding their effect on CMB can be drawn from our discussion. The power spectrum of the overall curvature perturbations $\zeta$ which are still superhorizon at least at matter-radiation equality can be written as:
\begin{equation}
\label{poweranisotropic}
P_{\zeta} ({\bf k})=P_{\zeta} (k) \left[1+A \cdot (\hat{{\bf B}} \cdot \hat{{\bf k}})^2 \right] \; ,
\end{equation}
where the amplitude $A$ is given by
\begin{equation}
A \simeq \frac{25 \cdot f^2}{324}  \cdot \frac{P_B (k, t_0)}{B^2_0 \cdot P_{\zeta} (k)} \;
\end{equation}
(see the end of the Appendix for the details of the derivation). That is, we deal with a quadrupolar directional-dependence~\cite{Ackerman:2007nb}. When PMF fluctuations are nearly flat, the amplitude $A$ is almost $k$-independent. The non-observation of statistical anisotropy with these properties in the WMAP~\cite{Hanson:2009gu, Ramazanov:2012za, Ramazanov:2013wea} and Planck datasets~\cite{Planck 2018_X, Kim:2013gka, Rubtsov:2014yua, Ramazanov:2016gjl} yields the upper limit $|A| \lesssim 0.02$. Notably, the values corresponding to the upper bound here can be easily reached in the model at hand, which means that the latter can be tested. To be more precise, in our case the formula~\eqref{poweranisotropic} correctly captures only the behaviour of very long wavelength fluctuations corresponding to low CMB multipoles, $l \lesssim 100$. Because of that the existing limit on the amplitude $A$ should be degraded by a factor of 10. Nevertheless, this fact will not play any profound role in our further discussion because the more constraining bound comes from the non-observation of isocurvature modes. Lastly, it is remarkable that the quantities $A$ and $P_{{\cal S}_{DM}}/P_{\zeta}$ are tightly connected to each other in the model at hand. Given also that the Planck data have a similar sensitivity to these quantities, it would be interesting to perform a simultaneous search for DM isocurvature perturbations and statistical anisotropy.

A direction-dependent power spectrum of scalar perturbations is of course not a signature unique to our model. For example, statistical anisotropy commonly appears in inflationary scenarios involving a coupling of (dark) electromagnetism to the inflaton such as $f(\phi) F^{\mu \nu} F_{\mu \nu}$, where $f(\phi)$ is typically an exponential function. This coupling is exactly the one typically employed for magnetogenesis, and it gives rise to a statistically anisotropic inflationary power spectrum~\cite{Watanabe:2009ct, Bartolo:2012sd}. Statistical anisotropy is also predicted in certain alternatives to inflation~\cite{Rubakov:2009np, Libanov:2010nk, Libanov:2011hh}. Contrary to our case, all these works deal with statistically anisotropic \emph{adiabatic} perturbations, whereas in our model we have statistically anisotropic \emph{isocurvature} perturbations, see also Refs.~\cite{Nakayama:2019rhg, Nakayama:2020rka}. Thus, we expect different features in the CMB temperature fluctuations.

Before we conclude this Section, let us briefly mention the possible effect of PMF helicity on CMB fluctuations. A non-zero helicity leads to an additional contribution to the power spectrum of the PMF described by the following tensor structure:
\begin{equation}
\langle B_i ({\bf k}) B_j ({\bf q}) \rangle \propto  \epsilon_{ijl}{k^l} \delta ({\bf k}+{\bf q}) \; .
\end{equation}
It is evident from the presence of the anti-symmetric tensor $\epsilon_{ijl}$ that the contribution of the helical component of the PMF to the fluctuations of $\chi$ vanishes. Hence, our mechanism is blind to any PMF helicity, which we can then neglect.

\section{Constraints on the parameter space}

In this Section we discuss the constraints on the model parameter space which follow from self-consistency of our assumptions, as well as consistency with observational data. In order not to run into contradiction with the non-observation of statistical anisotropy (with the caveats discussed in the previous Section) and DM isocurvature mode, one should impose the bound
\begin{equation}
f \lesssim \left(0.2-0.3\right) \cdot \sqrt{\frac{P_{\zeta} (k) \cdot B^2_0}{P_{B} (k, t_0)}} \; .
\end{equation}
In particular, if the inhomogeneity of the magnetic field is about $\delta B/B \simeq 10^{-5}$ as for radiation, then the parameter $f$ can be as large as $f \simeq 0.2-0.3$. On the other hand, there is no strong reason to assume that the magnetic fluctuations should be so small; indeed, they can be as large as $\delta B/B \simeq 1$. In that case, the fraction $f$ cannot exceed $f \simeq 10^{-6}$. Strictly speaking, it is possible to suppress the DM anisotropic isocurvature by assuming very small magnetic field fluctuations $\delta B/B \ll 10^{-5}$. 
If such a configuration is realised, then the PMF can be at the origin of the whole DM. In what follows, we keep $f$ as a free parameter, normalised to $0.1$ for concreteness.

Another important constraint follows from the condition $M \gg H_*$. With the use of Eq.~\eqref{massrB} this is interpreted as a limit on $\Lambda$:
\begin{equation}
\label{smalllambda}
\frac{\Lambda}{10^{14}~\mbox{GeV}} \ll 1.5 \cdot \left(\frac{g_* (T_*)}{10^3}\right)^{1/6} \cdot \left( \frac{r_B (t_0)}{10^{-10}} \right)^{1/2}  \; .
\end{equation}
This upper bound is close to the scale of Grand Unified Theory $\Lambda_{GUT} \simeq 10^{15}~\mbox{GeV}$ for $r_B (t_0) \simeq 10^{-10}$. Even though this might be a coincidence, we will see in what follows that the cutoff scale $\Lambda$ is confined to a narrow range adjacent to the upper limit~\eqref{smalllambda}.

Furthermore, the self-consistency of our scenario requires that higher order operators should be suppressed compared to that of Eq.~\eqref{nonrenorm}, at least during the inverse phase transition. In particular, this sets a bound on the value of the PMF: $B_* \ll \Lambda^2$. By virtue of Eq.~\eqref{relationPMF} this translates into $\Lambda \gg M$. The same limit follows from unitarity considerations: the annihilation cross-section of two particles $\chi$ into two photons is given by $\langle \sigma v \rangle_{\chi \chi \rightarrow \gamma \gamma} = M^2 / (2\pi \Lambda^4)$, see Refs.~\cite{Rajaraman:2012fu, Dudas:2014ixa}\footnote{The discrepancy by the factor $1/4$ compared to the expression given in Ref.~\cite{Dudas:2014ixa}, is due to the factor $1/2$ in Eq.~\eqref{nonrenorm}.}; the requirement that $\sigma E^2 \lesssim 1$ then gives $\Lambda \gg M$. In fact, the hierarchy between $\Lambda$ and $M$ is much stronger. Again, keeping the non-renormalisable operator~\eqref{nonrenorm} is possible only for small field values, in particular $\chi \ll \Lambda$. Then, barring any fine-tuning, in the spontaneously broken phase one should have $\chi \approx B/(\sqrt{\lambda}~ \Lambda)$. Combining the conditions $\chi \ll \Lambda$ and $B \gg M \Lambda$, we get
\begin{equation}
\label{limitstronger}
\Lambda \gg \frac{M}{\sqrt{\lambda}} \; .
\end{equation}
The limit has consequences for the transition temperature. Namely, from Eq.~\eqref{massrB}, one finds
\begin{equation}
\label{larger}
\frac{M}{T_*} \lesssim 5 \cdot 10^{-5} \left(\frac{g_* (T_*)}{10^3} \right)^{1/3} \cdot \left(\frac{\lambda}{10^{-11}} \right)^{1/4} \cdot \left(\frac{r_B (t_0)}{10^{-10}} \right)^{1/4}  \; .
\end{equation}
We conclude that the transition temperature $T_*$ is always larger than the mass $M$.

We also need to check the assumption that the inverse phase transition takes place at the radiation-dominated stage. Hence, the temperature $T_*$ cannot exceed the reheating temperature $T_{reh}$. This still leaves plenty of freedom, because $T_{reh}$ is allowed to vary in a very broad range between $4.2~\mbox{MeV}$~\cite{Hannestad:2004px} (close to the beginning of big-bang nucleosynthesis) and $10^{15}~\mbox{GeV}$. Using Eqs.~\eqref{massrB} and~\eqref{fraction}, from the inequality $T_* \lesssim T_{reh}$ we obtain:
\begin{equation}
\frac{M}{\mbox{TeV}} \lesssim \left(\frac{g_{*, reh}}{10^3} \right)^{1/5} \cdot \left(\frac{f}{0.1} \right)^{3/10} \cdot \left(\frac{\lambda}{10^{-11}} \right)^{3/10} \cdot \left(\frac{T_{reh}}{10^{10}~\mbox{GeV}} \right)^{1/2} \; ,
\end{equation}
or in terms of $\Lambda$: 
\begin{equation}
\label{upperreheating}
\frac{\Lambda}{10^{14}~\mbox{GeV}} \lesssim 0.7 \cdot \left(\frac{g_{*, reh}}{10^3} \right)^{7/15}  \cdot \left(\frac{0.1}{f} \right)^{3/10} \cdot \left(\frac{10^{-11}}{\lambda} \right)^{3/10} \cdot \left(\frac{r_B (t_0)}{10^{-10}}\right)^{1/2} \cdot \left(\frac{T_{reh}}{10^{10}~\mbox{GeV}} \right)^{3/2}\; . 
\end{equation}
From this point on, we assume $g_* (T_*) =g_{*, reh}$ for the sake of simplicity. Note that violating the latter two conditions does not rule out our scenario. This would simply mean that the inverse phase transition took place before the radiation-dominated stage, which means either during preheating or even during inflation, if the PMF existed at those early times--we do not discuss this option any further as it is beyond the scope of this work.

Up to now all constraints seem to allow for a relatively small $\Lambda \sim 10~\mbox{TeV}$. This could be good news from a point of view of testing this model with colliders or in astrophysical searches for $\chi$ particles. However, for relatively small $\Lambda$, the non-renormalisable interaction~\eqref{nonrenorm} leads to a fast thermalisation of $\chi$ with the hot plasma. As a result, the $\chi$-condensate produced through the inverse phase transition is rapidly destroyed, and the information about the PMF is drowned in the surrounding radiation. To avoid this we are forced to push $\Lambda$ towards the upper limit of Eq.~\eqref{smalllambda}, so that thermal equilibrium between the field $\chi$ and the primordial plasma is never attained. This sets a lower bound on the cutoff scale $\Lambda$. 

Thus, we require that the annihilation processes $\gamma +\gamma \leftrightarrow \chi +\chi$ as well as the scattering processes $\gamma +\chi \rightarrow \gamma +\chi$ involving $\gamma$ photons, are not efficient. In other words, one should fulfill the inequality:
\begin{equation}
\label{nonthermal}
\tau \sim \frac{1}{n_{\gamma, eq} \cdot \langle \sigma v\rangle } \gg H^{-1} \; ,
\end{equation}
where $\tau$ is the mean-free-path time of the photon prior to the collision with another photon (a particle $\chi$) leading to the creation of the pair of $\chi$ particles (scattering). The equilibrium number density of photons $n_{\gamma, eq}$ is given by
\begin{equation}
\label{photons}
n_{\gamma, eq} \approx \frac{2.4 \cdot T^3}{\pi^2} \; , 
\end{equation} 
and the cross-section $\langle \sigma v \rangle$ for the processes involving photons with energies $E \simeq T$ is estimated as~\cite{Rajaraman:2012fu, Dudas:2014ixa}
\begin{equation}
\label{cross}
\langle \sigma v \rangle  \simeq \frac{T^2}{2\pi \Lambda^4} \; . 
\end{equation}
The requirement that the inequality~\eqref{nonthermal} holds all the way back to the time of reheating sets the lower bound on $\Lambda$:
\begin{equation}
\label{pesky}
\frac{\Lambda}{10^{14}~\mbox{GeV}} \gg 3 \cdot 10^{-3} \cdot \left(\frac{10^3}{g_{*, reh}} \right)^{1/8} \cdot \left(\frac{T_{reh}}{10^{10}~\mbox{GeV}} \right)^{3/4}  \; .
\end{equation}

In fact, the lower bound on the cutoff scale $\Lambda$ should be further strengthened. The reason is that particles $\chi$ can be abundantly created by the freeze-in mechanism~\cite{McDonald:2001vt, Hall:2009bx}. Their production is governed by the equation
\begin{equation}
\label{freezein}
\frac{1}{a^3}\frac{d (n_{\chi} \cdot a^3)}{dt}=\langle \sigma v \rangle \cdot  (n^2_{\chi, eq}-n^2_{\chi}) \; .
\end{equation} 
We assume that the number density of $\chi$ particles, $n_{\chi}$, is negligible prior to reheating; furthermore, once the condition~\eqref{pesky} is fulfilled, we have $n_{\chi} \ll n_{\chi, eq}$, where $n_{\chi, eq} =\frac{1}{2} n_{\gamma, eq}$, and $n_{\gamma, eq}$ is given by Eq.~\eqref{photons}. The integral of Eq.~\eqref{freezein} is saturated at the lower bound, i.e., at the onset of the hot stage, when $T \simeq T_{reh}$:
\begin{equation}
n_{\chi} (t) \simeq \frac{1}{1.5\pi^5} \cdot \frac{T^8_{reh}}{\Lambda^4 \cdot H (T_{reh})} \cdot \left(\frac{a_{reh}}{a} \right)^3 \; .
\end{equation}
Therefore, the requirement that the energy density of particles $\chi$ produced by the freeze-in mechanism,
\begin{equation}
\rho_{\chi} = M n_{\chi} \; ,
\end{equation}
does not exceed the energy density of radiation at matter-radiation equality given by Eq.~\eqref{dmradeq}, gives the strong lower bound on $\Lambda$:
\begin{figure}[tb!]
  \begin{center}
    \includegraphics[width=0.8\columnwidth,angle=0]{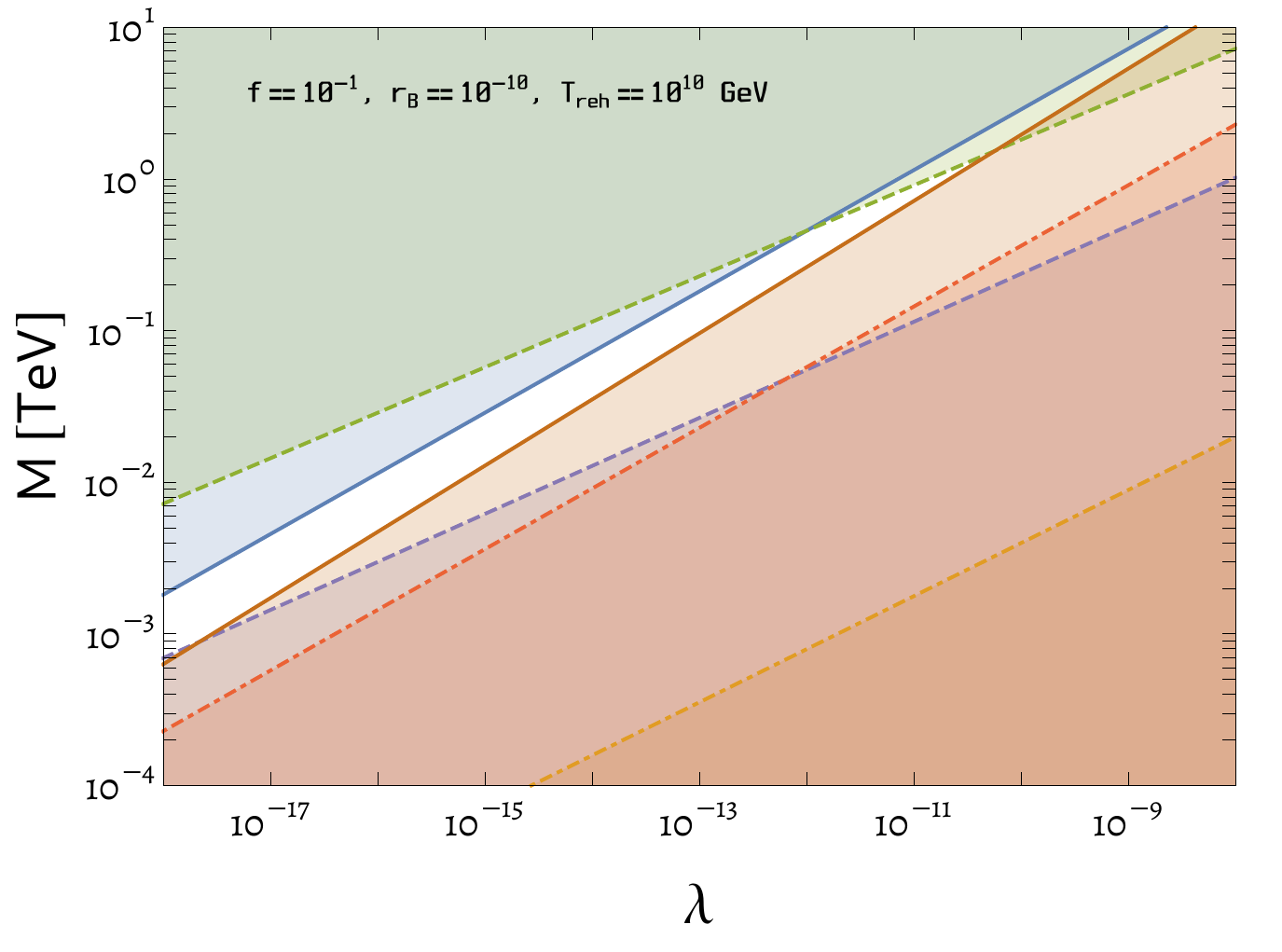}\\
    \includegraphics[width=\columnwidth, angle=0]{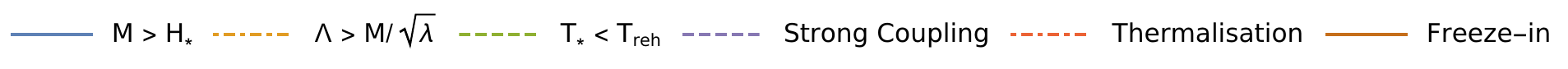}
  \caption{The exclusion plot shows allowed values (white region) for the mass $M$ of the field $\chi$ and the self-interaction coupling constant $\lambda$. The other parameters, the reheating temperature $T_{reh}$, the DM fraction $f$ defined by Eq.~\eqref{ft}, and the present day magnetic to radiation energy density ratio $r_B (t_0)$ defined by Eq.~\eqref{rB}, are set to the values shown in the plot. The number of ultra-relativistic degrees of freedom is fixed to $g_{*, reh}=g_* (T_*)=10^3$. For values of the mass $M$ and the coupling constant $\lambda$ along the solid orange line, the rest of DM is made of particles $\chi$ produced through the freeze-in mechanism.}\label{main1}
  \end{center}
\end{figure}

\begin{figure}[tb!]
  \begin{center}
    \includegraphics[width=0.8\columnwidth,angle=0]{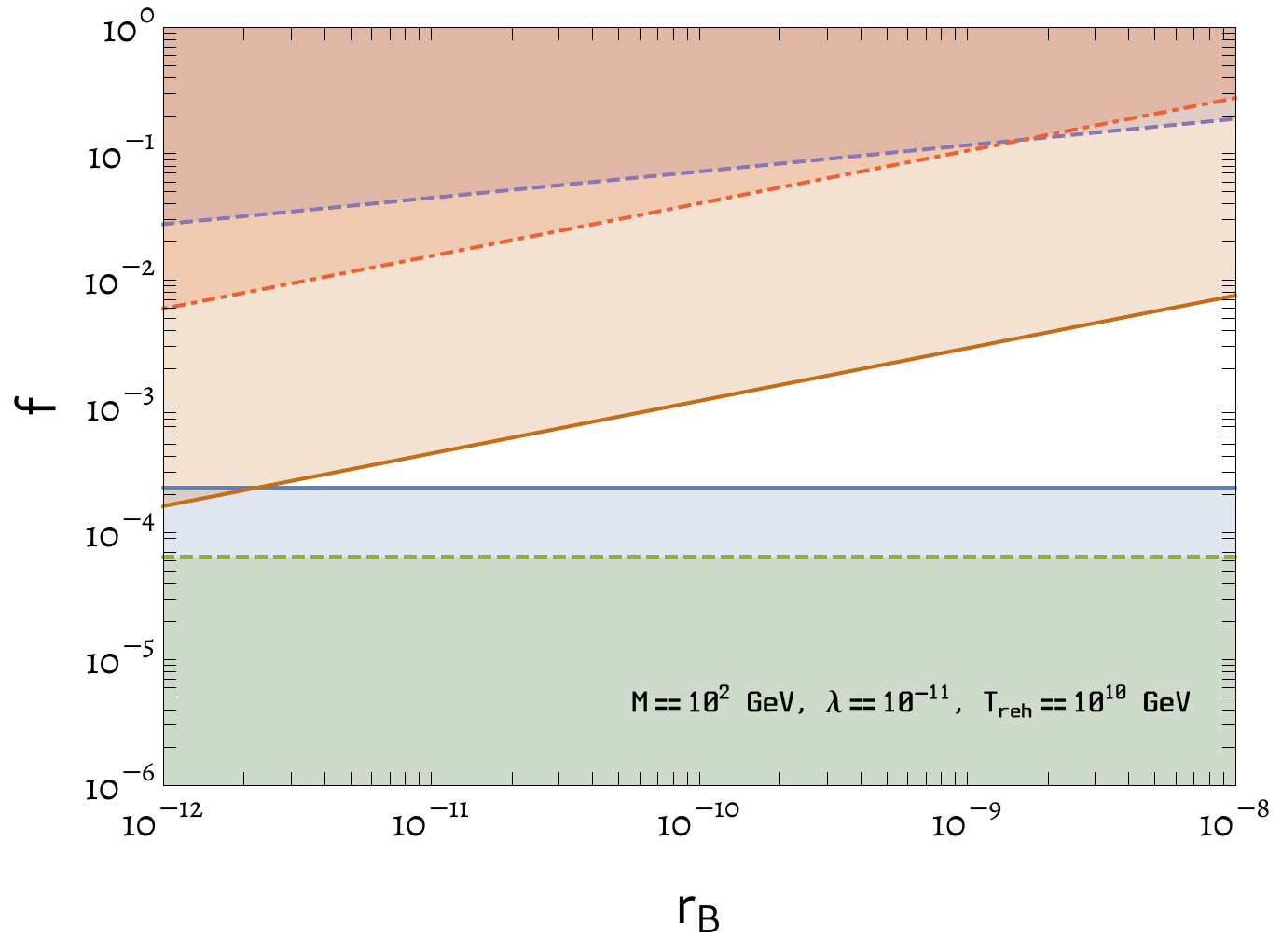}\\
        \includegraphics[width=\columnwidth, angle=0]{leg.pdf}
  \caption{The exclusion plot shows allowed values (white region) of the DM fraction $f$ defined by Eq.~\eqref{ft} and the present day magnetic to radiation energy density ratio $r_B (t_0)$ defined by Eq.~\eqref{rB}. The other parameters, the reheating temperature $T_{reh}$, the mass $M$ of the field $\chi$, and the self-interaction coupling constant $\lambda$, are set to the values shown in the plot. The number of ultra-relativistic degrees of freedom is fixed to $g_{*, reh}=g_* (T_*)=10^3$. For $f$ and $r_B (t_0)$ taking values along the solid orange line, the rest of DM is made of particles $\chi$ produced through the freeze-in mechanism.}\label{main2}
  \end{center}
\end{figure}

\begin{equation}
\label{lower}
 \frac{\Lambda}{10^{14}~\mbox{GeV}}  \gtrsim 0.4 \left(\frac{10^3}{g_{*, reh}} \right)^{3/8} \left(\frac{M}{\mbox{TeV}} \right)^{1/4}  \left(\frac{T_{reh}}{10^{10}~\mbox{GeV}} \right)^{3/4}  \; .
\end{equation}
One may wish to go to relatively low $\Lambda$ by lowering the reheating temperature. However, the consistency of the bounds~\eqref{upperreheating} and~\eqref{lower} implies that 
\begin{equation}
 \frac{T_{reh}}{10^{10}~\mbox{GeV}} \gtrsim 0.5 \cdot \left(\frac{10^3}{g_{*, reh}} \right)^{10/9} \cdot \left(\frac{f}{0.1} \right)^{2/5} \cdot \left(\frac{\lambda}{10^{-11}} \right)^{2/5} \cdot \left(\frac{10^{-10}}{r_B (t_0)} \right)^{2/3}  \cdot \left(\frac{M}{\mbox{TeV}} \right)^{1/3}\; .
\end{equation}
On the other hand, the consistency of Eqs.~\eqref{smalllambda} and~\eqref{lower} leads to an upper bound on the reheating temperature which reads:
\begin{equation}
\frac{T_{reh}}{10^{10}~\mbox{GeV}} \lesssim  6  \cdot \left(\frac{g_{*, reh}}{10^{3}} \right)^{13/18} \cdot \left(\frac{r_B (t_0)}{10^{-10}} \right)^{2/3} \cdot \left(\frac{\mbox{TeV}}{M} \right)^{1/3} \; .
\end{equation}

\begin{figure}[tb!]
  \begin{center}
    \includegraphics[width=\textwidth,angle=0,trim={1.5cm 1cm 1.5cm 1cm},clip]{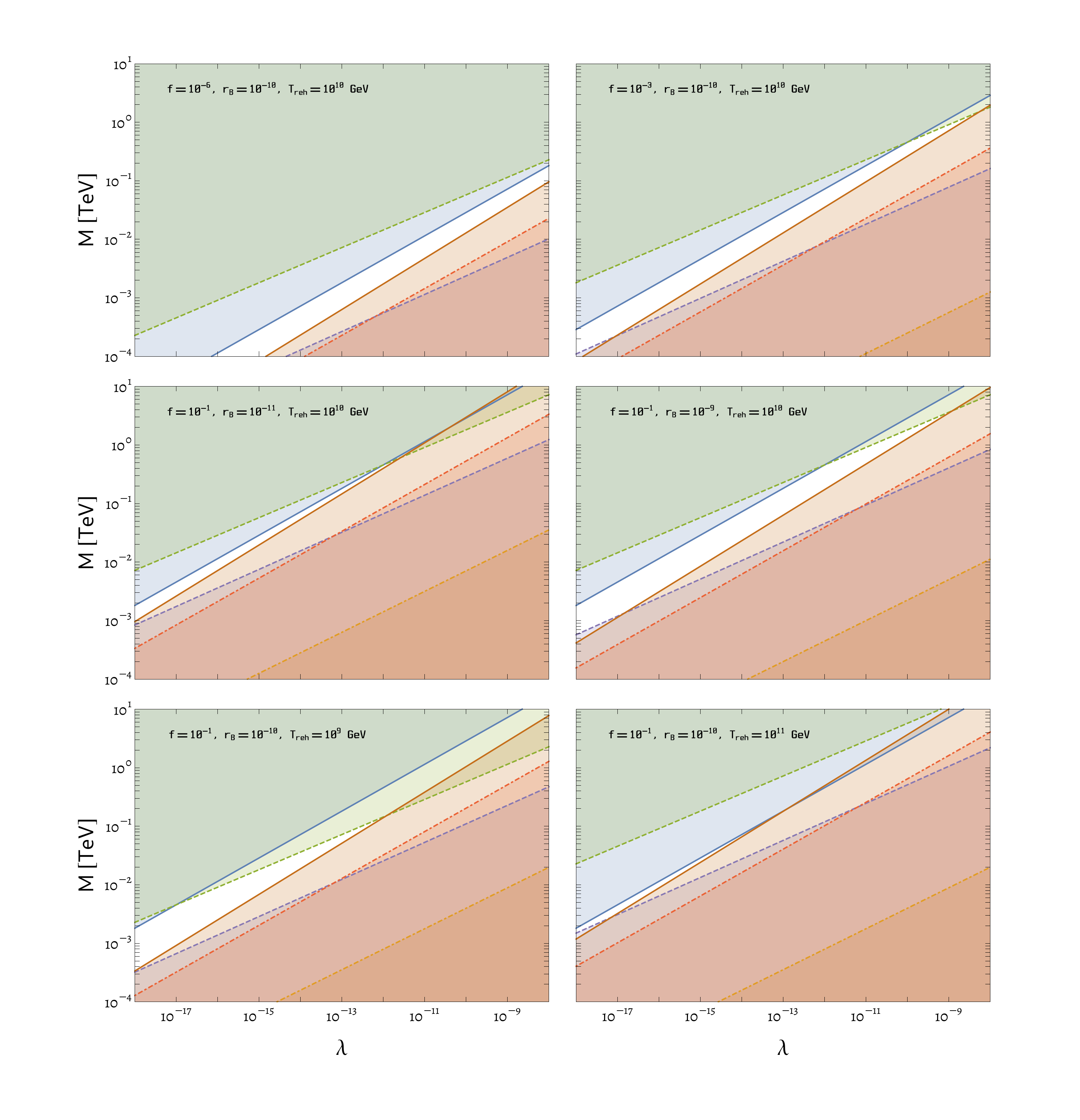}
    \caption{Exclusion plot for mass $M$ versus self-interaction coupling constant $\lambda$, as in Fig.~\ref{main1}, but for different DM fractions $f$ (top panel), present day magnetic to radiation energy density ratios $r_B (t_0)$ (middle panel), and reheating temperatures $T_{reh}$ (bottom panel).}\label{Mvarf}
  \end{center}
\end{figure}
 \begin{figure}[tb!]
  \begin{center}
    \includegraphics[width=\textwidth,angle=0,trim={1.5cm 1cm 1.5cm 1cm},clip]{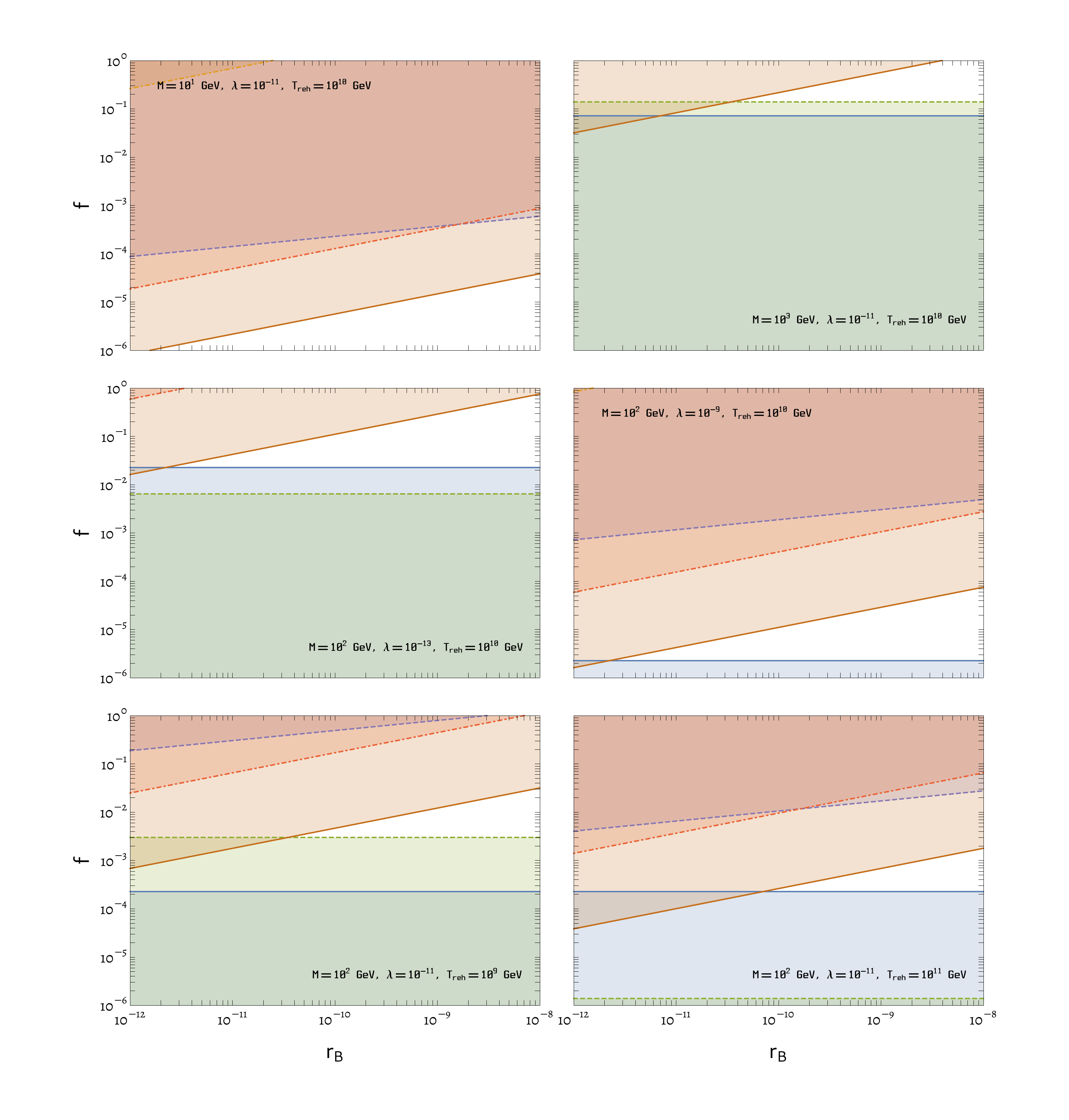}
    \caption{Exclusion plot for DM fraction $f$ versus present day magnetic to radiation energy density ratio $r_B (t_0)$, as in Fig.~\ref{main2}, but for different masses $M$ (top panel), self-interaction coupling constants $\lambda$ (middle panel), and reheating temperatures $T_{reh}$ (bottom panel).}\label{fvarM}
  \end{center}
\end{figure}

Two comments are in order here. At very high temperatures, the description of the SM gauge sector in terms of electromagnetism is not adequate, and hence the non-renormalisable interaction~\eqref{nonrenorm} should be UV-completed. However, working with the interaction~\eqref{nonrenorm} is legitimate, as long as it does not change the estimate of the cross-sections of the processes involving the particles $\chi$. Second, as it has been mentioned above, using the operator~\eqref{nonrenorm} is self-consistent only for field values $\chi \lesssim \Lambda$. For larger $\chi$ we enter the strong coupling regime, and one cannot trust the perturbative treatment. We require that the condition $\chi \lesssim \Lambda$ holds up to the beginning of the radiation-dominated stage (not only at the inverse phase transition). Otherwise, our estimate of the number density of $\chi$-particles produced by the freeze-in mechanism would not apply. Using Eqs.~\eqref{vacuum},~\eqref{relationPMF},~\eqref{rB}, and~\eqref{revolution}, one converts the inequality $\chi \lesssim \Lambda$ into the lower bound on the constant $\lambda$:
\begin{equation}
\lambda \gtrsim 5 \cdot 10^{-23} \cdot \left(\frac{r_B (t_0)}{10^{-10}} \right) \cdot \left(\frac{g_{*, reh}}{10^{3}} \right)^{4/3} \cdot \left(\frac{T_{reh}}{10^{10}~\mbox{GeV}} \right)^4 \cdot \left(\frac{10^{14}~\mbox{GeV}}{\Lambda} \right)^4 \; .
\end{equation}
As it will become clear below, this is a very weak bound for interesting $T_{reh}$ and $\Lambda$. 
 
Our final constraints on the parameter space are shown in Figs.~\ref{main1}, \ref{main2}, \ref{Mvarf}, and \ref{fvarM}. In particular, Figs.~\ref{main1} and \ref{Mvarf} and Figs.~\ref{main2} and \ref{fvarM} show projections 
of the parameter space on the planes $(M, \lambda)$ and $(f, r_B (t_0))$, respectively. We see that the strongest constraint comes from the requirement that the particles $\chi$ are not overproduced by the freeze-in mechanism. As a result, the parameter space turns out to be rather narrow. A positive outcome of this situation is that the particles $\chi$ created by the freeze-in mechanism naturally constitute most of DM. 

It is worth mentioning that the parameter space involves very small $r_B (t_0)$ down to $r_B (t_0) \simeq 10^{-12}$, see Figs.~\ref{main2} and \ref{fvarM}. Thus, our mechanism in principle allows to probe tiny PMF with the comoving amplitude $B_0 \simeq 10^{-3}\,\mbox{nG}$. Furthermore, the typical masses $M$ vary in the range $\mbox{GeV} \lesssim M \lesssim \mbox{TeV}$, which could be of interest from the viewpoint of ground based and astrophysical searches for DM. Finally, we would like to note that our parameter space favours tiny coupling constants $\lambda$. Despite this, a coupling constant $\lambda$ as large as $\lambda \simeq 10^{-6}$ is possible, provided that the DM fraction $f$ is very small, $f \simeq 10^{-6}$. Remarkably, such a tiny ``magnetomorphic'' DM fraction could still lead to observational consequences, as follows from the discussion in the beginning of this Section.  

\section{Conclusion}

In this work we have shown that even a very weak PMF may become visible in cosmological datasets when brought in contact with DM. We have demonstrated this by introducing a self-interacting scalar field $\chi$ coupled to electromagnetism through a Rayleigh non-renormalisable interaction $\chi^2 F_{\mu \nu} F^{\mu \nu}/\Lambda^2$. These interactions appear naturally in the low-energy limit of various extensions of SM. In some region of the parameter space such an interaction leads to an abundant production of $\chi$ particles through an ``inverse phase transition'', in which symmetry is restored as the Universe cools down. The properties of the PMF are then imprinted on the properties of the $\chi$ particles produced at the transition. It is then crucial that the latter are non-relativistic and sufficiently stable, so that they may comprise a considerable fraction of DM. Therefore, in this picture, by studying the properties of the ``magnetomorphic'' fraction of DM one could extract information about the PMF. This opens up the opportunity to probe a PMF with comoving amplitude $B \lesssim 0.01~\mbox{nG}$, which is otherwise inaccessible in current cosmological observations.

The most striking feature of this scenario is that, when PMF fluctuations have a flat spectrum, the scalar field $\chi$, acting as a DM component, carries \emph{statistically anisotropic isocurvature perturbations}. In particular, the detection of the latter would hint at the existence of long range vector modes in the primordial plasma coupled to the new stable particles $\chi$. In this way we could detect--albeit indirectly--a PMF and a sizeable fraction of DM at the same time. Because of the strict bounds on isocurvature modes as well as on statistical anisotropy in the CMB data, the particles $\chi$ created by the inverse phase transition may constitute only a part of DM; nonetheless, it is remarkable that the rest of DM can also be made of the same $\chi$ particles, this time produced via freeze-in.

The most stringent constraints on the parameter space of this model come from the requirement that one does not overproduce $\chi$ particles through the annihilation of $\gamma$-photons. On top of that, we have assumed that the inverse phase transition takes place during the radiation-dominated epoch; this further restricts the model parameters. Still, there is a non-zero parameter space even for a PMF as weak as $B_0 \simeq 0.01~\mbox{nG}$ which is compatible with all constraints and leads to observable signatures in near-future experiments. For example, one can choose $M=0.5~\mbox{TeV}$, $\Lambda =0. 4 \cdot 10^{14}~\mbox{GeV}$, and $\lambda=10^{-12}$ and assume the reheating temperature $T_{reh}=10^{10}~\mbox{GeV}$. Note that the constraints are considerably relaxed if the inverse phase transition occurs earlier, i.e., at preheating and even at inflation, which is a natural option if the PMF is generated by inflationary mechanisms.

It is possible to avoid some stringent constraints on the parameter space by enlarging the field content of our simple scenario. For example, one could consider the $\chi$-condensate non-perturbatively decaying into stable fermions $S$ with masses $M_S \lesssim M$ via a resonance akin to an inflaton. If such a decay is fast and efficient, one could have $\rho_{\chi} (t_*) \approx \rho_S (t_*)$. Hence, the field $S$ effectively takes on the role of $\chi$, and inherits the properties of the $\chi$-condensate. If stable enough, the particles $S$ contribute to DM. In this situation, the particles $\chi$ may have multiple channels of perturbative decay into SM species, so that the overproduction of $\chi$ particles through the annihilation of $\gamma$-photons can be avoided even for not very large $\Lambda$.

In the future, it would be interesting to investigate signatures of a PMF with a blue spectrum, such as those potentially arising from phase transitions. In that case, separating the PMF into a background part and a linear perturbation is not justified. Moreover, because the field $\chi$ interacts with the PMF quadratically as $\chi^2 B^2$, one expects a highly non-Gaussian imprint of PMF fluctuations on those of the field $\chi$. Thus, the latter, if stable on cosmological timescales, would constitute a non-Gaussian fraction of DM. If this fraction is large enough this non-Gaussianity could be detectable. Therefore, it may be possible to probe very weak PMF with a blue spectrum thanks to the indirect effects on DM.

\section*{Acknowledgments}

We are indebted to Camilo Garcia-Cely for many useful discussions. F.~U. and A.~V. are supported by the European Regional Development Fund (ESIF/ERDF) and the Czech Ministry of Education, Youth and Sports (M\v SMT) through the Project CoGraDS-CZ.02.1.01/0.0/0.0/15 003/0000437. A.~V. also acknowledges  support by the J. E. Purkyn\v{e} Fellowship of the Czech Academy of Sciences. The work of S.~R. is supported by the Czech Science Foundation GA\v CR, project 20-16531Y.

\section*{Appendix: Calculating Dark Matter isocurvature perturbation}

In this Appendix, we derive Eqs.~\eqref{finaliso} and~\eqref{poweranisotropic}. In the superhorizon regime, the equation of motion for the field $\chi$, which accounts for the fluctuations of the PMF, is given by
\begin{equation}
\label{quasihomo}
\ddot{\chi} (t, {\bf x})+3H \dot{\chi} (t, {\bf x})+M^2 \chi (t, {\bf x})+ \lambda \chi^3 (t, {\bf x})-\frac{{\bf B}^2 (t, {\bf x})}{\Lambda^2} \chi (t, {\bf x}) = 0 \; .
\end{equation}
We are interested in isocurvature modes excited by the PMF, so we have consistently set metric perturbations to zero. We work in the reference frame where the plasma is at rest and unperturbed. Eq.~\eqref{quasihomo} has the same form as the homogeneous equation, and hence the results of Section~2 are applicable. The important difference, however, is that the time when the inverse phase transition occurs depends on the position in space through the magnetic field ${\bf B} (t, {\bf x})$, which is slightly inhomogeneous. As a result, the energy density $\rho_{\chi}$ is defined as in the homogeneous case, but the amplitude $\chi_*$ and the beginning of oscillations $t_*$ must be promoted to functions of ${\bf x}$:
\begin{equation}
\rho_{\chi} (t, {\bf x})=\frac{M^2 \chi^2_* ({\bf x})}{2} \cdot \left(\frac{a (t_*({\bf x}))}{a(t)} \right)^3 \; .
\end{equation} 
Thus, the perturbation $\delta_{\chi, iso}$ can be expressed as
\begin{equation}
\label{start}
\delta_{\chi, iso}= \frac{2\delta \chi_* ({\bf x})}{\chi_*}+ 3 H_* \delta t_* ({\bf x}) \; .
\end{equation}
One finds $\delta \chi ({\bf x})/\chi$ by perturbing Eq.~\eqref{equality}:
\begin{equation}
\label{one} 
\frac{\delta \chi_* ({\bf x})}{\chi_*} =-\frac{2}{3} H_* \delta t_* ({\bf x}) +\frac{2}{3} \cdot \frac{\delta \mu_*}{\mu_*} \; .
\end{equation}
On the other hand, perturbing Eq.~\eqref{vacuum}, one obtains
\begin{equation}
\label{two} 
\frac{\delta \chi_* ({\bf x})}{\chi_*} \simeq \frac{\mu_* \delta \mu_* }{\lambda \chi^2_*} \; .
\end{equation}
We can then combine the expressions~\eqref{one} and~\eqref{two} taking into account that 
\begin{equation}
\frac{\lambda \chi^2_*}{\mu^2_{*}} \simeq \left(\frac{H_*}{M} \right)^{2/3} \ll 1 \; ; 
\end{equation} 
see Eqs.~\eqref{hierarchy} and~\eqref{maincase}. Hence,
\begin{equation}
\frac{\delta \mu_*}{\mu_*} \cdot \left[1- {\cal O} \left(\frac{\lambda \chi^2_*}{\mu^2_*} \right) \right] \simeq -\frac{2}{3} \frac{\lambda \chi^2_*}{\mu^2_*} \cdot H_* \delta t_* ({\bf x}) \; .
\end{equation}
We conclude that 
\begin{equation}
\label{zero}
\frac{\delta \mu_*}{\mu_*} \approx 0 \; .
\end{equation}
As it follows from Eqs.~\eqref{one} and~\eqref{zero}, the perturbation $\delta_{\chi, iso}$ given by Eq.~\eqref{start} can be written in the form
\begin{equation}
\label{isointer}
\delta_{\chi, iso} \approx \frac{5}{3} H_* \delta t_* ({\bf x}) \; .
\end{equation} 
Thus, to find $\delta_{\chi, iso}$ one should define $\delta t_* ({\bf x})$.

For this purpose we use Eq.~\eqref{zero}. Recall the connection~\eqref{relationPMF} between the value of the external field $\mu$ and the PMF. Hence,
\begin{equation}
\label{pert}
\frac{\delta \mu^2_*}{\mu^2_*} =\frac{\delta B^2 (t_* ({\bf x}), {\bf x})}{B^2_*}=\frac{2{\bf B}_* \delta {\bf B}_* (t_* ({\bf x}), {\bf x})}{B^2_*} \; .
\end{equation}
The perturbation $\delta {\bf B} (t_* ({\bf x}), {\bf x})$ has two sources: the difference between the values of the background field ${\bf B}$ following from the spatial dependence of the transition time $t_{*} ({\bf x})$, and 
the proper perturbation of the magnetic field itself $\delta {\bf B}_{*} ({\bf x})$ considered at the time $t_*$, 
\begin{equation}
\label{split}
\delta {\bf B} (t_* ({\bf x}), {\bf x}) = \mathbf{B}\left(t_{*}+\delta t_{*}\left(\mathbf{x}\right)\right)-\mathbf{B}\left(t_{*}\right)+\delta {\bf B}_* ({\bf x}) \; .
\end{equation}
We can then write $\mathbf{B}\left(t_{*}+\delta t_{*}\left(\mathbf{x}\right)\right)-\mathbf{B}\left(t_{*}\right)=-2H_* {\bf B}_* \delta t_* ({\bf x})$; upon substituting this into Eq.~\eqref{split} and the latter into Eq.~\eqref{pert}, we get 
\begin{equation}
\delta t_* ({\bf x})  \approx \frac{{\bf B}_* \cdot \delta {\bf B}_* ( {\bf x})}{2 H_* \cdot B^2_*} \; .
\end{equation}    
Finally, from Eq.~\eqref{isointer} we derive the desired expression for the perturbation $\delta_{\chi, iso}$, 
\begin{equation}
\delta_{\chi, iso}=\frac{5}{6} \cdot \frac{{\bf B}_{*} \cdot \delta {\bf B}_* ({\bf x})}{B^2_*} \; ,
\end{equation}
which we quoted in Eq.~\eqref{finaliso}.

Now, let us comment on the details of derivation of Eq.~\eqref{poweranisotropic}. For this purpose, we need the definition of the DM isocurvature perturbation given by Eq.~\eqref{iso}. Each $\zeta_i$ entering this definition remains nearly constant in the superhorizon regime, provided that the anisotropic stress is negligible (as is the case here). The same is not true for the total curvature perturbation, which changes beyond the horizon in presence of the isocurvature mode. The expression for the total curvature perturbation in terms of the individual curvature perturbations is given by
\begin{equation}
\label{zetatotal} 
\zeta=\frac{4/3 \rho_{rad} \cdot \zeta_{rad} +\rho_{DM} \cdot \zeta_{DM}}{4/3 \rho_{rad}+\rho_{DM}} \; 
\end{equation}
(for simplicity, we omitted the contribution due to baryons). At the matter-dominated stage one finds from Eqs.~\eqref{iso} and~\eqref{zetatotal}
\begin{equation}
\label{finin}
\zeta=\zeta_{rad}+\frac{1}{3} {\cal S}_{DM} \; ,
\end{equation}
see Ref.~\cite{Wands:2000dp}. Now, combining this with Eqs.~\eqref{isochi} and~\eqref{finaliso}, we get Eq.~\eqref{poweranisotropic}.

\clearpage


\begin{thebibliography}{99}

\bibitem{Durrer:2013pga}
  R.~Durrer and A.~Neronov,
  Astron.\ Astrophys.\ Rev.\  {\bf 21} (2013) 62
  [arXiv:1303.7121 [astro-ph.CO]].
  
\bibitem{Subramanian:2015lua}
  K.~Subramanian,
  Rept.\ Prog.\ Phys.\  {\bf 79} (2016) no.7,  076901
  [arXiv:1504.02311 [astro-ph.CO]].
  
\bibitem{Giovannini:2003yn}
  M.~Giovannini,
  Int.\ J.\ Mod.\ Phys.\ D {\bf 13} (2004) 391
  [astro-ph/0312614].

\bibitem{Beck:2000dc}
R.~Beck,
Space Sci. Rev. \textbf{99} (2001), 243-260
[arXiv:astro-ph/0012402 [astro-ph]].

\bibitem{Bernet:2008qp}
  M.~L.~Bernet, F.~Miniati, S.~J.~Lilly, P.~P.~Kronberg and M.~Dessauges-Zavadsky,
  Nature {\bf 454} (2008) 302
  [arXiv:0807.3347 [astro-ph]].
  
\bibitem{Carilli:2001hj}
C.~L.~Carilli and G.~B.~Taylor,
Ann. Rev. Astron. Astrophys. \textbf{40} (2002), 319-348
[arXiv:astro-ph/0110655 [astro-ph]].

\bibitem{Govoni:2004as}
F.~Govoni and L.~Feretti,
Int. J. Mod. Phys. D \textbf{13} (2004), 1549-1594
[arXiv:astro-ph/0410182 [astro-ph]].

\bibitem{Widrow:2002ud}
L.~M.~Widrow,
Rev. Mod. Phys. \textbf{74} (2002), 775-823
[arXiv:astro-ph/0207240 [astro-ph]].


 
\bibitem{Kulsrud:2007an}
R.~M.~Kulsrud and E.~G.~Zweibel,
Rept. Prog. Phys. \textbf{71} (2008), 0046091
[arXiv:0707.2783 [astro-ph]].
 
\bibitem{Brandenburg:2004jv}
A.~Brandenburg and K.~Subramanian,
Phys. Rept. \textbf{417} (2005), 1-209
[arXiv:astro-ph/0405052 [astro-ph]].

\bibitem{Dolag:2002bw}
K.~Dolag, M.~Bartelmann and H.~Lesch,
Astron. Astrophys. \textbf{387} (1999), 386
[arXiv:astro-ph/0202272 [astro-ph]].

\bibitem{Dolag:2004kp}
  K.~Dolag, D.~Grasso, V.~Springel and I.~Tkachev,
  JCAP {\bf 0501} (2005) 009
  [astro-ph/0410419].
    
\bibitem{Turner:1987bw}
  M.~S.~Turner and L.~M.~Widrow,
  Phys.\ Rev.\ D {\bf 37} (1988) 2743.
  
\bibitem{Ratra:1991bn}
  B.~Ratra,
  Astrophys.\ J.\ Lett.\  {\bf 391} (1992) L1.
  
\bibitem{Martin:2007ue}
  J.~Martin and J.~Yokoyama,
  JCAP {\bf 0801} (2008) 025
  [arXiv:0711.4307 [astro-ph]].
  
\bibitem{Demozzi:2009fu}
  V.~Demozzi, V.~Mukhanov and H.~Rubinstein,
  JCAP {\bf 0908} (2009) 025
  [arXiv:0907.1030 [astro-ph.CO]].
  
\bibitem{Kanno:2009ei}
  S.~Kanno, J.~Soda and M.~a.~Watanabe,
  JCAP {\bf 0912} (2009) 009
  [arXiv:0908.3509 [astro-ph.CO]].
  

\bibitem{Urban:2011bu}
F.~R.~Urban,
JCAP \textbf{12} (2011), 012
[arXiv:1111.1006 [astro-ph.CO]].

\bibitem{Emami:2009vd}
R.~Emami, H.~Firouzjahi and M.~S.~Movahed,
Phys. Rev. D \textbf{81} (2010), 083526
[arXiv:0908.4161 [hep-th]].
  
\bibitem{Ferreira:2013sqa}
R.~J.~Z.~Ferreira, R.~K.~Jain and M.~S.~Sloth,
JCAP \textbf{10} (2013), 004
[arXiv:1305.7151 [astro-ph.CO]].



\bibitem{Kobayashi:2014sga}
T.~Kobayashi,
JCAP \textbf{05} (2014), 040
[arXiv:1403.5168 [astro-ph.CO]].

\bibitem{Fujita:2016qab}
T.~Fujita and R.~Namba,
Phys. Rev. D \textbf{94} (2016) no.4, 043523
[arXiv:1602.05673 [astro-ph.CO]].

\bibitem{Vilchinskii:2017qul}
S.~Vilchinskii, O.~Sobol, E.~Gorbar and I.~Rudenok,
Phys. Rev. D \textbf{95} (2017) no.8, 083509
[arXiv:1702.02774 [astro-ph.CO]].

\bibitem{Fujita:2015iga}
T.~Fujita, R.~Namba, Y.~Tada, N.~Takeda and H.~Tashiro,
JCAP \textbf{05} (2015), 054
[arXiv:1503.05802 [astro-ph.CO]].

\bibitem{Adshead:2016iae}
P.~Adshead, J.~T.~Giblin, T.~R.~Scully and E.~I.~Sfakianakis,
JCAP \textbf{10} (2016), 039
[arXiv:1606.08474 [astro-ph.CO]].

\bibitem{Caprini:2014mja}
C.~Caprini and L.~Sorbo,
JCAP \textbf{10} (2014), 056
[arXiv:1407.2809 [astro-ph.CO]].



\bibitem{Sharma:2018kgs}
R.~Sharma, K.~Subramanian and T.~R.~Seshadri,
Phys. Rev. D \textbf{97} (2018) no.8, 083503
[arXiv:1802.04847 [astro-ph.CO]].

\bibitem{Fujita:2019pmi}
T.~Fujita and R.~Durrer,
JCAP \textbf{09} (2019), 008
[arXiv:1904.11428 [astro-ph.CO]].

\bibitem{Domenech:2015zzi}
G.~Domenech, C.~Lin and M.~Sasaki,
EPL \textbf{115} (2016) no.1, 19001
[arXiv:1512.01108 [astro-ph.CO]].

\bibitem{Mukohyama:2016npi}
S.~Mukohyama,
Phys. Rev. D \textbf{94} (2016) no.12, 121302
[arXiv:1607.07041 [hep-th]].



\bibitem{Talebian:2020drj}
A.~Talebian, A.~Nassiri-Rad and H.~Firouzjahi,
[arXiv:2007.11066 [gr-qc]].

\bibitem{Patel:2019isj}
T.~Patel, H.~Tashiro and Y.~Urakawa,
JCAP \textbf{01} (2020), 043
[arXiv:1909.00288 [astro-ph.CO]].


\bibitem{Battefeld:2004cd}
T.~J.~Battefeld and R.~Brandenberger,
Phys. Rev. D \textbf{70} (2004), 121302
[arXiv:hep-th/0406180 [hep-th]].

\bibitem{Salim:2006nw}
J.~M.~Salim, N.~Souza, S.~E.~Perez Bergliaffa and T.~Prokopec,
JCAP \textbf{04} (2007), 011
[arXiv:astro-ph/0612281 [astro-ph]].

\bibitem{Sriramkumar:2015yza}
L.~Sriramkumar, K.~Atmjeet and R.~K.~Jain,
JCAP \textbf{09} (2015), 010
[arXiv:1504.06853 [astro-ph.CO]].

\bibitem{Chowdhury:2016aet}
D.~Chowdhury, L.~Sriramkumar and R.~K.~Jain,
Phys. Rev. D \textbf{94} (2016) no.8, 083512
[arXiv:1604.02143 [gr-qc]].

\bibitem{Qian:2016lbf}
P.~Qian, Y.~F.~Cai, D.~A.~Easson and Z.~K.~Guo,
Phys. Rev. D \textbf{94} (2016) no.8, 083524
[arXiv:1607.06578 [gr-qc]].

\bibitem{Chowdhury:2018blx}
D.~Chowdhury, L.~Sriramkumar and M.~Kamionkowski,
JCAP \textbf{01} (2019), 048
[arXiv:1807.05530 [astro-ph.CO]].

\bibitem{Frion:2020bxc}
E.~Frion, N.~Pinto-Neto, S.~D.~P.~Vitenti and S.~E.~Perez Bergliaffa,
Phys. Rev. D \textbf{101} (2020) no.10, 103503
[arXiv:2004.07269 [gr-qc]].

  
\bibitem{Vachaspati:1991nm}
  T.~Vachaspati,
  Phys.\ Lett.\ B {\bf 265} (1991) 258.
  
\bibitem{Enqvist:1993np}
K.~Enqvist and P.~Olesen,
Phys. Lett. B \textbf{319} (1993), 178-185
[arXiv:hep-ph/9308270 [hep-ph]].
  


  
\bibitem{Joyce:1997uy}
  M.~Joyce and M.~E.~Shaposhnikov,
  Phys.\ Rev.\ Lett.\  {\bf 79} (1997) 1193
  [astro-ph/9703005].
  
\bibitem{Durrer:2003ja}
  R.~Durrer and C.~Caprini,
  JCAP {\bf 0311} (2003) 010
  [astro-ph/0305059].
  
\bibitem{Kahniashvili:2012uj}
  T.~Kahniashvili, A.~G.~Tevzadze, A.~Brandenburg and A.~Neronov,
  Phys.\ Rev.\ D {\bf 87} (2013) no.8,  083007
  [arXiv:1212.0596 [astro-ph.CO]].
  
  
\bibitem{Jedamzik:1996wp}
  K.~Jedamzik, V.~Katalinic and A.~V.~Olinto,
  Phys.\ Rev.\ D {\bf 57} (1998) 3264
  [astro-ph/9606080].
  
\bibitem{Subramanian:1997gi}
  K.~Subramanian and J.~D.~Barrow,
  Phys.\ Rev.\ D {\bf 58} (1998) 083502
  [astro-ph/9712083].

  
\bibitem{Neronov:1900zz}
  A.~Neronov and I.~Vovk,
  Science {\bf 328} (2010) 73
  [arXiv:1006.3504 [astro-ph.HE]].
  



\bibitem{Jedamzik:2018itu}
  K.~Jedamzik and A.~Saveliev,
  Phys.\ Rev.\ Lett.\  {\bf 123} (2019) no.2,  021301
  [arXiv:1804.06115 [astro-ph.CO]].
  
\bibitem{Jedamzik:2020krr}
  K.~Jedamzik and L.~Pogosian,
  arXiv:2004.09487 [astro-ph.CO].
  
\bibitem{Ade:2015cva}
  P.~A.~R.~Ade {\it et al.} [Planck Collaboration],
  Astron.\ Astrophys.\  {\bf 594} (2016) A19
  [arXiv:1502.01594 [astro-ph.CO]].

\bibitem{Trivedi:2013wqa}
P.~Trivedi, K.~Subramanian and T.~R.~Seshadri,
Phys. Rev. D \textbf{89} (2014) no.4, 043523
[arXiv:1312.5308 [astro-ph.CO]].

\bibitem{Blasi:1999hu}
P.~Blasi, S.~Burles and A.~V.~Olinto,
Astrophys. J. Lett. \textbf{514} (1999), L79-L82
[arXiv:astro-ph/9812487 [astro-ph]].
  
\bibitem{Pshirkov:2015tua}
  M.~S.~Pshirkov, P.~G.~Tinyakov and F.~R.~Urban,
  Phys.\ Rev.\ Lett.\  {\bf 116} (2016) no.19,  191302
  [arXiv:1504.06546 [astro-ph.CO]].


\bibitem{Nugaev:2020zcv}
  E.~Nugaev and A.~Shkerin,
  arXiv:2004.14354 [hep-ph].

\bibitem{Babichev:2020xeg}
  E.~Babichev, D.~Gorbunov and S.~Ramazanov,
  JCAP {\bf 2008} (2020) 047
  [arXiv:2004.03410 [hep-ph]].

\bibitem{Laulumaa:2020pqi}
  L.~Laulumaa, T.~Markkanen and S.~Nurmi,
  JCAP {\bf 2008} (2020) 002
  [arXiv:2005.04061 [astro-ph.CO]].
  
\bibitem{Bettoni:2018pbl}
D.~Bettoni, G.~Dom\`enech and J.~Rubio,
JCAP \textbf{02} (2019), 034
[arXiv:1810.11117 [astro-ph.CO]].

  
  
\bibitem{Hollenstein:2012mb}
L.~Hollenstein, R.~K.~Jain and F.~R.~Urban,
JCAP \textbf{01} (2013), 013
[arXiv:1208.6547 [astro-ph.CO]].

\bibitem{Nakayama:2019rhg}
K.~Nakayama,
JCAP \textbf{10} (2019), 019
[arXiv:1907.06243 [hep-ph]].

\bibitem{Nakayama:2020rka}
K.~Nakayama,
JCAP \textbf{08} (2020), 033
[arXiv:2004.10036 [hep-ph]].

  
\bibitem{Chang:2010en}
S.~Chang, N.~Weiner and I.~Yavin,
Phys. Rev. D \textbf{82} (2010), 125011
[arXiv:1007.4200 [hep-ph]].

\bibitem{Weiner:2012gm}
N.~Weiner and I.~Yavin,
Phys. Rev. D \textbf{87} (2013) no.2, 023523
[arXiv:1209.1093 [hep-ph]].

\bibitem{Liu:2013gba}
J.~Liu, B.~Shuve, N.~Weiner and I.~Yavin,
JHEP \textbf{07} (2013), 144
[arXiv:1303.4404 [hep-ph]].

\bibitem{Rajaraman:2012fu}
A.~Rajaraman, T.~Tait, M.P. and A.~M.~Wijangco,
Phys. Dark Univ. \textbf{2} (2013), 17-21
[arXiv:1211.7061 [hep-ph]].


\bibitem{Dudas:2014ixa}
  E.~Dudas, L.~Heurtier and Y.~Mambrini,
  Phys.\ Rev.\ D {\bf 90} (2014) 035002
  [arXiv:1404.1927 [hep-ph]].
  
   \bibitem{Feinberg1970}
    G.~Feinberg, J.~Sucher,
    Phys. \ Rev. \ A {\bf 2} (1970) 2395--2415
    
    
  \bibitem{Itzyk}
  C.~Itzykson, J.~B.~Zuber, 
  "Quantum Field Theory",
  McGraw-Hill (1980) 

\bibitem{Adamek:2011pr}
  J.~Adamek, R.~Durrer, E.~Fenu and M.~Vonlanthen,
  JCAP {\bf 1106} (2011) 017
  [arXiv:1102.5235 [astro-ph.CO]].

\bibitem{Giovannini:2007qn}
  M.~Giovannini and K.~E.~Kunze,
  Phys.\ Rev.\ D {\bf 77} (2008) 063003
  [arXiv:0712.3483 [astro-ph]].

\bibitem{Yamazaki:2008gr}
  D.~G.~Yamazaki, K.~Ichiki, T.~Kajino and G.~J.~Mathews,
  Phys.\ Rev.\ D {\bf 77} (2008) 043005
  [arXiv:0801.2572 [astro-ph]].

\bibitem{Paoletti:2008ck}
  D.~Paoletti, F.~Finelli and F.~Paci,
  Mon.\ Not.\ Roy.\ Astron.\ Soc.\  {\bf 396} (2009) 523
  [arXiv:0811.0230 [astro-ph]].

\bibitem{Bonvin:2011dt}
  C.~Bonvin, C.~Caprini and R.~Durrer,
  Phys.\ Rev.\ D {\bf 86} (2012) 023519
  [arXiv:1112.3901 [astro-ph.CO]].
  
\bibitem{Bonvin:2013tba}
  C.~Bonvin, C.~Caprini and R.~Durrer,
  Phys.\ Rev.\ D {\bf 88} (2013) 083515
  [arXiv:1308.3348 [astro-ph.CO]].


    
\bibitem{Planck 2018_X}
Planck Collaboration, Y. Akrami \textit{et al.}. 
Astron.~Astrophys. \textbf{641} (2020) A10,  
[arXiv: 1807.06211 [astro-ph.CO]]

\bibitem{Malik:2002jb}
K.~A.~Malik, D.~Wands and C.~Ungarelli,
Phys. Rev. D \textbf{67} (2003), 063516,
[arXiv:astro-ph/0211602 [astro-ph]].

\bibitem{Kodama:1985bj}
H.~Kodama and M.~Sasaki,
Prog. Theor. Phys. Suppl. \textbf{78} (1984), 1-166

  
\bibitem{Ackerman:2007nb}
  L.~Ackerman, S.~M.~Carroll and M.~B.~Wise,
  Phys.\ Rev.\ D {\bf 75} (2007) 083502
   Erratum: [Phys.\ Rev.\ D {\bf 80} (2009) 069901]
  [astro-ph/0701357].
 
\bibitem{Hanson:2009gu}
  D.~Hanson and A.~Lewis,
  Phys.\ Rev.\ D {\bf 80} (2009) 063004
  [arXiv:0908.0963 [astro-ph.CO]].
  
\bibitem{Ramazanov:2012za}
  S.~R.~Ramazanov and G.~I.~Rubtsov,
  JCAP {\bf 1205} (2012) 033
  [arXiv:1202.4357 [astro-ph.CO]].
  
\bibitem{Ramazanov:2013wea}
  S.~R.~Ramazanov and G.~Rubtsov,
  Phys.\ Rev.\ D {\bf 89} (2014) no.4,  043517
  [arXiv:1311.3272 [astro-ph.CO]].

    
\bibitem{Kim:2013gka}
  J.~Kim and E.~Komatsu,
  Phys.\ Rev.\ D {\bf 88} (2013) 101301
  [arXiv:1310.1605 [astro-ph.CO]].
  
\bibitem{Rubtsov:2014yua}
  G.~I.~Rubtsov and S.~R.~Ramazanov,
  Phys.\ Rev.\ D {\bf 91} (2015) no.4,  043514
  [arXiv:1406.7722 [astro-ph.CO]].
  
\bibitem{Ramazanov:2016gjl}
  S.~Ramazanov, G.~Rubtsov, M.~Thorsrud and F.~R.~Urban,
  JCAP {\bf 1703} (2017) 039
  [arXiv:1612.02347 [astro-ph.CO]].
  
  
  
\bibitem{Watanabe:2009ct}
  M.~a.~Watanabe, S.~Kanno and J.~Soda,
  Phys.\ Rev.\ Lett.\  {\bf 102} (2009) 191302
  [arXiv:0902.2833 [hep-th]].

\bibitem{Bartolo:2012sd}
  N.~Bartolo, S.~Matarrese, M.~Peloso and A.~Ricciardone,
  Phys.\ Rev.\ D {\bf 87} (2013) no.2,  023504
  [arXiv:1210.3257 [astro-ph.CO]].

\bibitem{Thorsrud:2013mma}
M.~Thorsrud, D.~F.~Mota and F.~R.~Urban,
Phys. Lett. B \textbf{733} (2014), 140-143
doi:10.1016/j.physletb.2014.04.028
[arXiv:1311.3302 [astro-ph.CO]].

\bibitem{Thorsrud:2013kya}
M.~Thorsrud, F.~R.~Urban and D.~F.~Mota,
JCAP \textbf{04} (2014), 010
doi:10.1088/1475-7516/2014/04/010
[arXiv:1312.7491 [astro-ph.CO]].

\bibitem{Rubakov:2009np}
  V.~A.~Rubakov,
  JCAP {\bf 0909} (2009) 030
  [arXiv:0906.3693 [hep-th]].
  
\bibitem{Libanov:2010nk}
  M.~Libanov and V.~Rubakov,
  JCAP {\bf 1011} (2010) 045
  [arXiv:1007.4949 [hep-th]].

\bibitem{Libanov:2011hh}
  M.~Libanov, S.~Ramazanov and V.~Rubakov,
  JCAP {\bf 1106} (2011) 010
  [arXiv:1102.1390 [hep-th]].

\bibitem{Hannestad:2004px}
  S.~Hannestad,
  Phys.\ Rev.\ D {\bf 70} (2004) 043506
  [astro-ph/0403291].

\bibitem{McDonald:2001vt}
J.~McDonald,
Phys. Rev. Lett. \textbf{88} (2002), 091304
[arXiv:hep-ph/0106249 [hep-ph]].

  
\bibitem{Hall:2009bx}
L.~J.~Hall, K.~Jedamzik, J.~March-Russell and S.~M.~West,
JHEP \textbf{03} (2010), 080
[arXiv:0911.1120 [hep-ph]].

\bibitem{Wands:2000dp}
  D.~Wands, K.~A.~Malik, D.~H.~Lyth and A.~R.~Liddle,
  Phys.\ Rev.\ D {\bf 62} (2000) 043527
  [astro-ph/0003278].
  

\end{thebibliography}
\end{document}